\documentclass[aps,pre,twocolumn,english,showpacs,amsmath,amssymb,superscriptaddress,longbibliography]{revtex4-1}
\usepackage{color}
\usepackage{mathptmx}

\usepackage[T1]{fontenc}
\usepackage{float}
\usepackage{graphicx}
\usepackage{xcolor}
\usepackage[normalem]{ulem}
\usepackage{soul}
\usepackage[latin9]{inputenc}
\usepackage{amsmath}
\usepackage{amssymb}
\usepackage{stackrel}
\usepackage{bm}
\usepackage{color}
\usepackage{framed,color}
\definecolor{shadecolor}{gray}{0.85}
\usepackage{comment}
\usepackage{multirow}

\makeatletter

\newcommand{\Rmnum}[1]{\expandafter\@slowromancap\romannumeral #1@}

\makeatother

\usepackage{babel}

\usepackage{bm}
\usepackage{xr}
\begin{document}

	\title{Distinct signature of two local structural motifs of liquid water\\ in the scattering function}

	\author{Rui Shi}
	\affiliation{Department of Fundamental Engineering, Institute of Industrial Science, \\
		University of Tokyo, 4-6-1 Komaba, Meguro-ku, Tokyo 153-8505, Japan}
	\author{Hajime Tanaka}
	\email{tanaka@iis.u-tokyo.ac.jp}
	\affiliation{Department of Fundamental Engineering, Institute of Industrial Science, \\
		University of Tokyo, 4-6-1 Komaba, Meguro-ku, Tokyo 153-8505, Japan}
	
	\date{\today}
	
	\begin{abstract}
		{Liquids generally become more ordered upon cooling. However, it has been a long-standing debate on whether such structural ordering in liquid water takes place continuously or discontinuosly: continuum vs. mixture models. Here, by computer simulations of three popular water models and analysis of recent scattering experiment data, we show that, in the structure factor of water, there are two overlapped peaks hidden in the apparent ``first diffraction peak'', one of which corresponds to the neighboring O-O distance as in ordinary liquids and the other to the longest periodicity of density waves in a tetrahedral structure. This unambiguously proves the coexistence of two local structural motifs. Our findings not only provide key clues to settle long-standing controversy on the water structure but also allow experimental access to the degree and range of structural ordering in liquid water.
		}
	\end{abstract}

	\maketitle

Water is ubiquitous in our planet and plays vital roles in many biological, geological, meteorological, and technological processes. Despite its simple molecular structure, water shows many unique thermodynamic and dynamic properties in the liquid state, such as a density maximum at 4~$^\circ$C, a rapid increase of isothermal compressibility and a dynamic fragile-to-strong transition upon cooling~\cite{Debenedetti2003,gallo2016water}. These unusual properties, which are absent in ordinary liquids, are well-known as ``water's anomalies''. It is widely believed that the anomalies are linked to water's structural ordering towards tetrahedral structures stabilized by four hydrogen bonds (H-bonds). Even after intensive studies for more than a century, however, how such structural ordering takes place is still a matter of hot debate without convergence. 

Two conflicting different scenarios have continued to exist until now: `continuum models' based on a broad unimodal distribution of structural components and `mixture models' based on a bimodal distribution of structural components reflecting the coexistence of two (or more) types of local structures~\cite{narten1969observed,eisenberg2005structure,handle2017supercooled}. 
The mixture model dates back to Wilhelm R\"ontgen, who proposed in 1892 that water can be regarded as 'icebergs' in a fluid 'sea'~\cite{rontgen1892ueber}. Later various mixture models have been developed. One famous example is the mixture model of Linus Pauling, who proposed in 1959 that water is mixture of clathrate-like structure and interstitial molecules~\cite{pauling1959structure}. These mixture models, however, have been continuously challenged by the continuum model dating back to John Pople, who proposed in 1951 that water's structure can be described by a continuously distorted H-bond network~\cite{pople1951molecular}. 

These two types of models lead to fundamentally different understandings of water structure. Despite such a clear difference in the physical picture, there has been no convergence of this debate over a century. 
The main reason is the lack of {\it experimental} evidence exclusively supporting either of the two models. In this Letter, we provide such clear evidence that liquid water is indeed a mixture of two types of local structural motifs, from simulations of three popular water models and detailed analysis of recent scattering experiments.

First we need to explain the precise nature of our two-state model to clarify essential  
differences from `continuum models' and other types of mixture models that regard water as a mixture of two types of liquids, i.e. low-density (LDL) and high-density liquids (HDL). 
Our two-state model that regards water as a mixture of ordered ($S$) and less ordered local structural motifs ($\rho$-state)~\cite{Tanaka_review,Tanaka2000} is characterized by the following five features: 
(1) $S$- and $\rho$-states in liquid water are defined by {\it local structures around a central molecule} and characterized by low and high local symmetry, energy, density, and entropy, respectively. In the one-phase regime of water far from the second-critical point (if it exists), the two structural motifs can have only short coherence lengths. Thus, our model is essentially different from a mixture model of LDL and HDL. We stress that they are  macroscopic phases of water that can exist only below the second critical point;
(2) Reflecting the presence of the two states, the distribution of a proper structural descriptor should have a bimodal distribution composed of two Gaussians  (not necessarily two delta functions; note that there is no  unique configuration for each state under thermal fluctuations); (3) The two-state model effectively transforms to a continuum-like model at high temperatures/high pressures where there exists only $\rho$ state, because the ordered $S$-structure can hardly survive due to the entropy/volume penalties; (4) The $T, P$-dependence of the fraction of the two structural motifs should obey the thermodynamic two-state equations~\cite{Tanaka_review}.  (5) The existence of a second critical point is a sufficient but not necessary condition for the two-state model.

\begin{center}
	\begin{figure}[b!]
		\includegraphics[width=14.5cm]{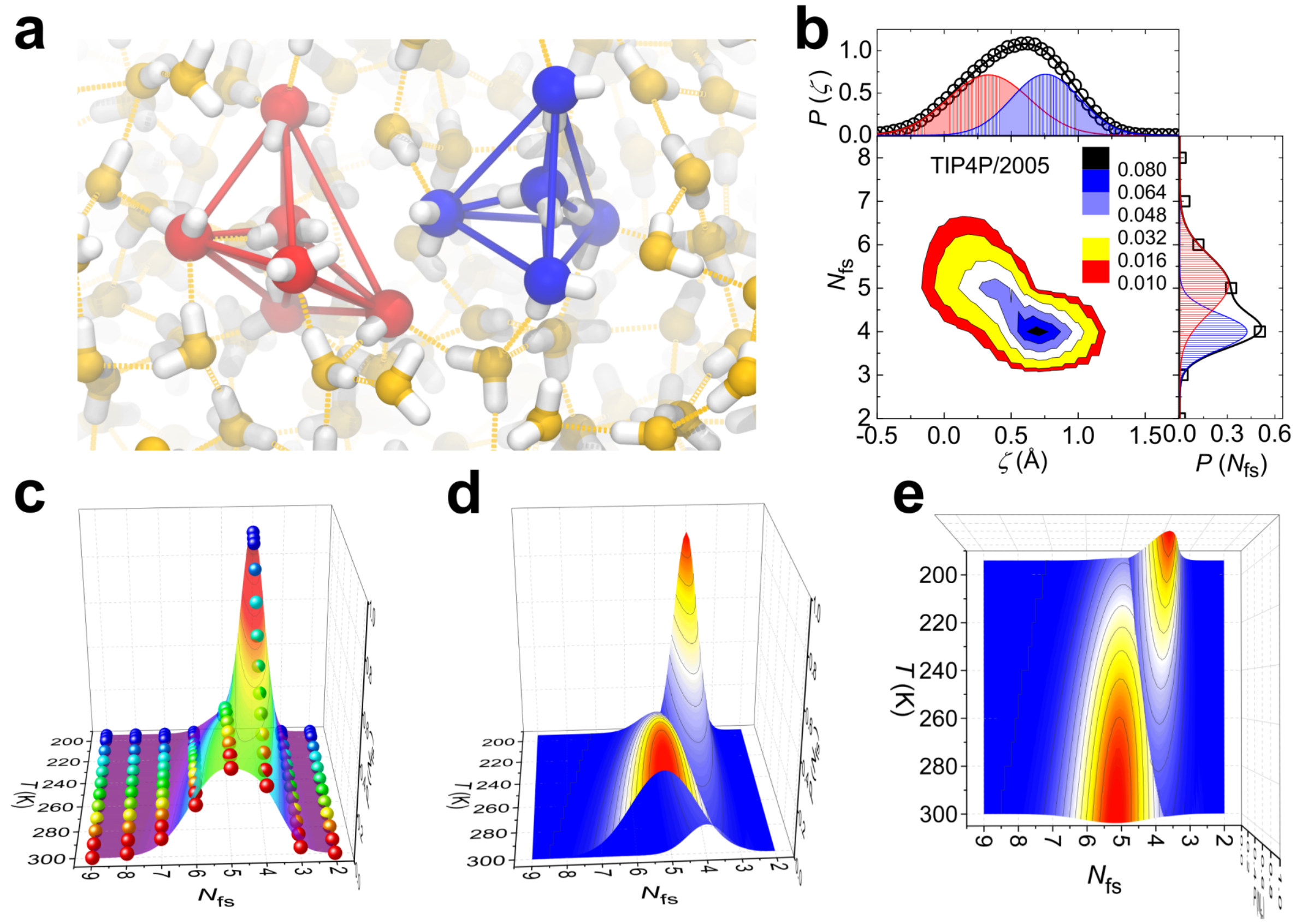}
	\vspace{-4mm}
	\caption{Structural bimodality in the coordination number distribution of liquid water. {\bf a}, A snapshot of liquid TIP4P/2005 water at 1 bar and 240 K, where $s\approx 0.5$, i.e., near the Schottky temperature $T_{s=1/2}$. Two structural motifs, LFTS and DNLS, are highlighted by blue and red colors, respectively. LFTS has four H-bonded nearest neighbors with tetrahedral coordination, whereas DNLS typically has five or more nearest neighbors (typically three of them being H-bonded) with broken tetrahedral symmetry. {\bf b}, Correlation between the number of water molecules in the spherical first-shell volume of radius of 3.5~\AA, $N_\mathrm{fs}$  (i.e., the coordination number), and the structural descriptor $\zeta$ at 1 bar and 240 K. The distributions of $N_\mathrm{fs}$ and $\zeta$ are shown in the right and top sides, respectively. Both distributions show clear bimodal features, which can be properly described by the sum of two Gaussian functions (black lines), with blue and red shades corresponding to LFTS and DNLS respectively. {\bf c}, Distribution of coordination number $N_\mathrm{fs}$ as a function of temperature, $P \left(N_\mathrm{fs}, T\right)$, is shown by colored balls with red and blue for higher and lower temperatures, respectively. The colored surface is the fit to two Gaussian functions by Eqs.~\ref{eq:pnfs} - \ref{eq:SDNLS}. {\bf d}-{\bf e}, $T$-dependence of the two Gaussian components of $P \left(N_\mathrm{fs}, T\right)$ in side ({\bf d}) and top ({\bf e}) view. One Gaussian component locating at $N_\mathrm{fs} \approx 4$ corresponds to LFTS, whereas the other one at $N_\mathrm{fs} \simeq 5$ corresponds to DNLS, in agreement with the snapshot in {\bf a}. The fraction of the two Gaussian components is consistent with the fraction determined by $\zeta$ distribution in {\bf b} and the theoretical two-state model (Eq.~\ref{eq:s}).}
	\label{fig:fnfsZeta}
\end{figure}
\end{center}

Recently we have shown that we can detect two structure motifs by a microscopic structural descriptor $\zeta$ (see Methods), which characterizes the translational order in the second shell, and confirmed the above five features on a microscopic level by computer simulations of several popular water models~\cite{Russo2014,shi2018impact,shi2018Microscopic,shi2018origin,shi2018common}. 
These studies have clearly indicated that water is a dynamic mixture of the two states~\cite{Tanaka_review,Tanaka2000,holten2013nature,Russo2014,singh2016two,Biddle2017,
Singh2017,de2018viscosity,russo2018water,shi2018impact,shi2018common,shi2018origin}---$S$-state [locally favored tetrahedral structure (LFTS)] and $\rho$-state [disordered normal-liquid structure (DNLS)]. The former stabilized by four H-bonds has lower symmetry, density, energy and entropy than the latter. A typical snapshot of LFTS and DNLS is shown in Fig.~\ref{fig:fnfsZeta}a. We have also found that the fraction $s$ of the LFTS, serving as an order parameter characterizing the degree of structural ordering, changes with temperature $T$ and pressure $P$, obeying the prediction of the thermodynamic two-state model~\cite{Tanaka_review,Tanaka2000,Singh2017,de2018viscosity,russo2018water,shi2018common,shi2018origin}. 
These results provide strong computational support for the two-state model.

The shape of  the distribution function of a physical quantity is a key to judge which of mixture and continuum models is relevant, because the former predicts a bimodal distribution at a certain range of $T$ and $P$ whereas the latter always predicts a unimodal Gaussian distribution.  
Our structural descriptor $\zeta$ clearly shows the bimodality. Then, a key question is whether 
a quantity directly related to local density shows such bimodality or not. The answer is yes. 
We show the distribution $P(N_\mathrm{fs})$ of the coordination number $N_\mathrm{fs}$ in Fig.~\ref{fig:fnfsZeta}b. 
We note that $N_\mathrm{fs}$ is the number of water molecules in the spherical first-shell volume $V_{\rm fs}$ of radius of 3.5~\AA, and thus proportional to the local number density, $N_{\rm fs}/V_{\rm fs}$ 
(see Methods (Characterization of local density) for the relevance of this estimation of local density). We can see that $P(N_\mathrm{fs})$ has clear bimodality. Furthermore, $P(N_\mathrm{fs})$ and $P(\zeta)$ both can be properly characterized by two Gaussian functions (see Methods) with the same fraction $s$ (see Fig.~\ref{fig:fnfsZeta}b), following the prediction of the thermodynamic two-state model (Fig.~\ref{fig:fnfsZeta}c-e and Fig.~\ref{fig:fnfsT}-\ref{fig:fnfs2Gau-small}). 
This clearly indicates the anti-correlation between $\zeta$ and local density [see the above feature (1)]. 
This result strongly contradicts with the prediction of the continuum models that $P(N_\mathrm{fs})$ should be unimodal Gaussian, which is the case for simple liquids such as Lennard-Jones liquids (Fig.~\ref{fig:fnfs-lj}). We note that three popular water models all show the bimodal distributions of $P(N_\mathrm{fs})$ (Fig.~\ref{fig:fnfsT}). 
We can see that $P(N_\mathrm{fs})$ exhibits a unimodal distribution at very high $T$, but it transforms to a bimodal one (composed of two Gaussians) upon cooling for all the three water models. 
This clearly indicates the failure of continuum models, and supports the two-state model (see the above Features (1)-(4)).

\onecolumngrid
\vspace{-6mm}
\twocolumngrid

It has sometimes been argued that the unimodal Gaussian distribution of density fluctuations is a signature against mixture models. However, we point out that it is not the case: Under thermal fluctuations, any  thermodynamic order parameters, e.g. density $\rho$ and local structural order $s$, should have unimodal Gaussian distributions. This is because the free energy of a system, $f(\rho,s)$, can be expressed by quadratic terms in  the one-phase homogeneous region (see, e.g., Ref.~\cite{tanaka1998simple}).  
Although theoretically obvious, we can confirm it from the fact that the macroscopic density distributions in liquid water and other single- or two-component liquids commonly show Gaussian distributions (see the results of  Lennard-Jones (LJ) liquid, SiO$_2$, and Cu$_{64}$Zr$_{36}$ in Fig.~\ref{fig:fdensity}) irrespective of whether 
the local density distribution is unimodal or bimodal (unimodal for LJ liquid whereas bimodal for H$_2$O, SiO$_2$ and Cu$_{64}$Zr$_{36}$). The same is applied for the distribution of another macroscopic order parameter $s$ (estimated from $\zeta$): $P(s)$ has a unimodal Gaussian distribution, even when the underlying microscopic structural descriptor $\zeta$ has a bimodal distribution composed of two Gaussians (Fig.~\ref{fig:fs_zeta}). This difference between macroscopic and microscopic distributions clearly indicates that the bimodal structural ordering in liquid water is highly localized, in agreement with Feature (1) in the introduction. Here we note that a mixture model of LDL and HDL should result in the bimodal distributions of macroscopic order parameters $\rho$ and $s$, contrary to the above results.

So far we show that computer simulations of classical water models allow us to directly access the distributions of $\zeta$ and $N_\mathrm{fs}$ and provide strong evidence for the presence of the two types of structural motifs. Unfortunately, however, we cannot access such microscopic molecular-level information by experiments. So an experimentally accessible structural descriptor is highly desirable to close a long-standing debate on the structure of liquid water.


The most powerful experimental methods to access the local structures of materials are x-ray and neutron scatterings, by which we can measure the structure factor, i.e. the density-density correlation in reciprocal space:  
\begin{equation}
S(\bm{k}) = \frac{1}{N}\langle\rho_{\bm{k}} \rho_{-\bm{k}}\rangle
\label{eq:sk}
\end{equation}
where $\langle \cdots \rangle$ denotes the ensemble average, $N$ is the number of particles, $\rho_{\bm{k}} = \sum_{i=1}^{N} \exp \left(-i\bm{k}\cdot \bm{r}_{i}\right)$ is the number density, $\bm{r}_i$ is the position vector of particle $i$, and $\bm{k}$ is the wave vector. In a crystal, the density $\rho_{\bm{k}}$ has components only at particular wave vectors $\bm{k}$'s because of the periodic arrangement of particles, leading to sharp diffraction spots at those wave vectors in the structure factor. These spots provide a complete description of a crystal structure. On the other hand, liquids and amorphous solids do not possess long-range translational order, and, as a result, only board isotropic amorphous halos are usually observed, which makes their structural characterization extremely difficult. This has also been the case for liquid water. So far no evidence of the coexistence of two types of structural motifs has been detected in $S(k)$ ($k=|\bm k|$), which has been a main cause of a continuous doubt on the two-state model. In this Letter, however, we report a new analysis of $S(k)$ focusing on the first few peaks, which provides direct experimental evidence for the coexistence of the two types of structural motifs and thus supports the two-state model.

\begin{center}
	\begin{figure}[b!]
		\includegraphics[width=16cm]{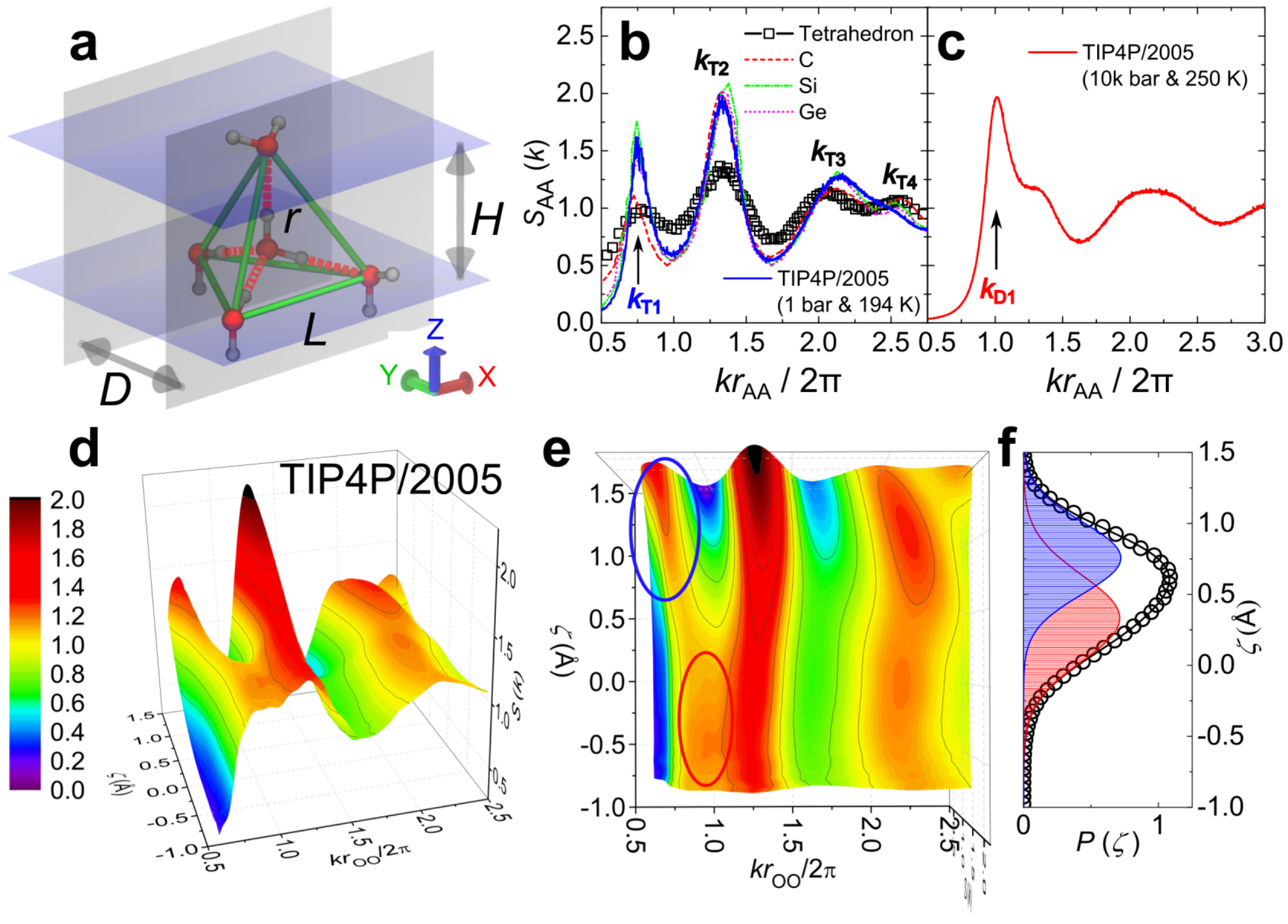}
		\vspace{-3mm}
		\caption{Structural bimodality in the structure factor of liquid water. {\bf a}, A schematic representation of a regular tetrahedron formed by five water molecules with the nearest O-O distance $r_{\rm OO}$, height $H$, width $D$, edge length $L$. {\bf b}, The O-O partial structure factor $S_\mathrm{AA}(k)$ ($\mathrm{AA}=\mathrm{OO}$) of a regular tetrahedron (formed by five oxygen atoms) and simulated TIP4P/2005 liquid water at 1 bar, 194 K, together with the structure factor of typical amorphous tetrahedral materials, C~\cite{gilkes1995}, Si~\cite{laaziri1999} and Ge~\cite{etherington1982}. Deeply supercooled liquid water clearly shows four characteristic peaks ($k_{Ti}$ ($i=1 \sim 4$)) common to tetrahedral materials. The peak position of FSDP, $k_{T1}=kr_\mathrm{OO}/2\pi \simeq 3/4 $, indicated by the arrow cooresponds to the height $H$ of an LFTS~\cite{shi2019distinct}. {\bf c}, The O-O partial structure factor of TIP5P/2005 water at 10000 bar, 250 K (see Fig.~\ref{fig:sqooP} for results of real and TIP5P water). It shows a characteristic peak of normal disordered systems at $k_{D1}=kr_\mathrm{OO}/2\pi \simeq 1 $ (see the arrow). {\bf d},{\bf e}, The $\zeta$-dependent partial O-O structure factor $S(k,\zeta)$ of TIP4P/2005 water at 1 bar, 240 K, calculated by Debye's scattering equation (Eqs.~\ref{eq:ski} - \ref{eq:skzeta}), in side ({\bf d}) and top ({\bf e}) view (see Fig.~\ref{fig:sqooZeta} for TIP5P and ST2 models). The characteristic peaks, $k_\mathrm{T1}$ and $k_\mathrm{D1}$, are highlighted by blue and red circles, respectively. {\bf f}, The distribution of $\zeta$ shows two Gaussian components, corresponding to LFTS (blue shade) and DNLS (red shade) respectively. The wave numbers in {\bf b}, {\bf c}, {\bf d}, {\bf e} are scaled by the nearest neighbor distance $r_\mathrm{AA}$ for all cases (A = C, Si, Ge and O in water).}
		\label{fig:sqooTetraZeta}
	\end{figure}
\end{center}

To do so, we focus on the lowest wave number peak in liquid water. In simple liquids such as hard spheres and LJ liquids, the first diffraction peak usually appears at the wave number $k \hat{r}/2\pi \simeq 1$ corresponding to the average nearest neighbor distance $\hat{r}$, or the average interparticle distance. 
However, it has been reported that $S(k)$ of a wide class of materials has a peak at a lower wave number whose corresponding length is longer than the average nearest neighbor distance~\cite{elliott1991medium}. Such a peak has been widely observed in the so-called tetrahedral liquids such as SiO$_2$, GeO$_2$, BeF$_2$, Si, Ge and C, and widely known as the first sharp diffraction peak (FSDP)~\cite{shi2019distinct}. 
The emergence of FSDP has been considered as a signature of intermediate-range structural ordering in liquids and amorphous states. Recently we have discovered~\cite{shi2019distinct} that FSDP of these liquids originates from the scattering from the density wave characteristic of a tetrahedral unit in LFTS, which is the fundamental structural motif of tetrahedral materials. More precisely, a density wave whose wave vector corresponds to the height $H$ of the tetrahedral structure (e.g., along the $Z$ direction in Fig.~\ref{fig:sqooTetraZeta}a) generates a sharp diffraction peak specifically at $k_\mathrm{T1}=k \hat{r}/2\pi \simeq 3/4$, i.e., FSDP. 
We note that a tetrahedral unit produces four peaks in the range $0.5 \leq k \hat{r}/2\pi \leq 3.0$ with peak wave numbers labelled as $k_{\mathrm{T}i}$ ($i=1\sim4$) from low to high $k$~\cite{shi2019distinct} (Fig.~\ref{fig:sqooTetraZeta}b).
If the two structural motifs revealed by $\zeta$ for water models are also relevant to real water, there should be the corresponding distinct signatures in the experimentally measured structure factor. Such a signature is indeed seen from the locations of the first diffracton peak in the structure factor of low $T$ and high $P$ water (Fig.~\ref{fig:sqoo_lowT_highP}). We can see a more distinct signature in simulated model waters, for which we are able to access both much lower temperatures (predominantly composed of LFTS), and higher pressures (predominantly composed of DNLS) than for experiments, without suffering from ice crystallisation. Figures~\ref{fig:sqooTetraZeta}b and c show the partial O-O structure factor of TIP4P/2005 water at low $T$ (LFTS-dominant) and high $P$ (DNLS-dominant), respectively, together with those of typical amorphous tetrahedral materials, C, Si and Ge. We can clearly see that low-$T$ water shows the structure factor very similar to the typical amorphous tetrahedral materials and its FSDP is exactly located at $k_\mathrm{T1}=k \hat{r}/2\pi \simeq 3/4$, as expected~\cite{shi2019distinct}. On the other hand, high-$P$ water has a first diffraction peak at $k_\mathrm{D1}=k \hat{r}/2\pi \simeq 1$, as simple liquids do, reflecting its (partially) disordered nature. In the two-state regime lying between the two extreme conditions, where LFTS and DNLS coexist with comparable populations, distinct signatures from the two structural motifs are expected to appear in the structure factor of liquid water.
 

To reveal local structural characteristics in the wave-number space, we employ what is called ``the Debye scattering function''~\cite{debye1915zerstreuung} (see Eqs.~\ref{eq:skdebye} - \ref{eq:skzeta} and Fig.~\ref{fig:sqdebye}). This allows us to access the correlation between the local structure of each structural motif characterized by $\zeta$ and its local structure factor on the firm theoretical basis. 
Figures~\ref{fig:sqooTetraZeta}d and e show the $\zeta$ dependent O-O structure factor $S(k,\zeta)$ of TIP4P/2005 water at 1 bar and 240 K, where water has equal amount of LFTS and DNLS (or, $s=1/2$). We refer this particular temperature to the Schottky temperature~\cite{shi2018common} and denote it as $T_{s=1/2}$.  Strikingly, we can see two distinct peaks at $k_\mathrm{T1}$ and $k_\mathrm{D1}$ in different $\zeta$ domains, which are nicely characterized by the two Gaussian components in the distribution of $\zeta$. 
Thus, we may conclude that the two peaks at $k_\mathrm{T1}$ and $k_\mathrm{D1}$ in the structure factor should correspond to LFTS and DNLS, respectively (Figs.~\ref{fig:sqooTetraZeta}e and f). We have also confirmed the same feature for TIP5P and ST2 water (Fig.~\ref{fig:sqooZeta}). Together with the bimodality of the structural descriptor $\zeta$ and coordination number $N_\mathrm{fs}$, this result further supports the two-state model. 
We emphasize that our finding indicates that we can now access the two-state signature {\it experimentally} by analysing the structure factor of real water.

Unfortunately, however, because the $k_\mathrm{T1}$ and $k_\mathrm{D1}$ peaks are close to each other, they are heavily overlapped under substantial thermal fluctuations, which makes a clear separation difficult. 
This difficulty is a source of the long-standing controversy. 
Thanks to the strong two-state nature in liquid silica---a tetrahedral liquid structurally similar to water~\cite{shi2018impact}---and large scattering cross sections of the atoms, we recently found that the apparent ``first diffraction peak'' in the Si-Si partial structure factor of silica is indeed a doublet: A Lorentzian peaked at $k_\mathrm{T1}$ and a Gaussian peaked at $k_\mathrm{D1}$ are necessary to properly describe the apparent ``first diffraction peak''. Moreover, the integrated intensity of the $k_\mathrm{T1}$ component is proportional to the fraction $s$ of LFTS, which is determined independently from a microscopic structural descriptor $z$~\cite{shi2018impact}; namely, it obeys the prediction of the thermodynamic two-state model~\cite{shi2019distinct}.

\onecolumngrid
\twocolumngrid

\onecolumngrid
\begin{center}
\begin{figure}[h!]
		\includegraphics[width=16.6cm]{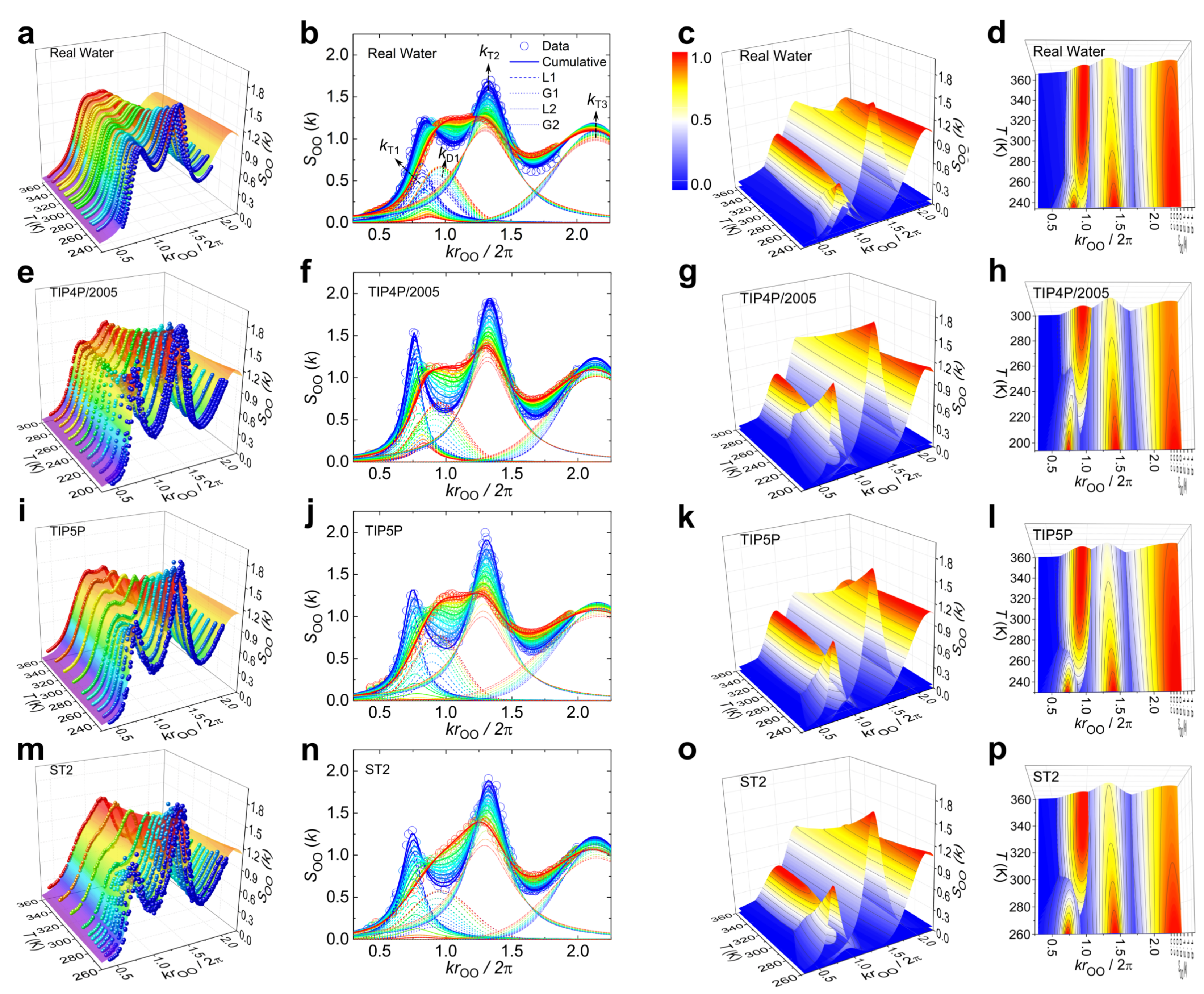}
	\vspace{-0mm}
	\caption{Analysis of O-O partial structure factors of real water and model waters at ambient pressure. {\bf a}-{\bf d}, X-ray scattering data of real water~\cite{skinner2014structure,pathak2019intermediate}. {\bf e}-{\bf h}, TIP4P/2005 model. {\bf i}-{\bf l}, TIP5P model. {\bf m}-{\bf p}, ST2 model. The O-O partial structure factors are shown by spheres in 3D plots ({\bf a}, {\bf e}, {\bf i}, {\bf m}) and circles in 2D plots ({\bf b}, {\bf f}, {\bf j}, {\bf n}) with more blue and red color for lower and higher temperatures respectively. The colored surfaces in ({\bf a}, {\bf e}, {\bf i}, {\bf m}) are the fits of our model (two Lorentzian (L1 + L2) and two Gaussian (G1 + G2) functions; see Eq.~\ref{eq:sk2} for the details) to the structure factors. In {\bf b}, {\bf f}, {\bf j}, {\bf n}, four characteristic peaks obtained from the fit are displayed by broken lines and assigned as indicated by the arrows in {\bf b}. The colored surfaces in the right two columns show the temperature dependences of the four characteristic peaks from the fit in side view ({\bf c}, {\bf g}, {\bf k}, {\bf o}) and top view ({\bf d}, {\bf h}, {\bf l}, {\bf p}). The color bar is shown in {\bf c} and the number in it denotes the ratio of the peak height to the maximum height of each peak over the temperature and wave number ranges shown in each image. The wave number is scaled by the nearest neighbor O-O distance $r_\mathrm{OO}$.}
	\label{fig:sqooT}
\end{figure}
\end{center}
\twocolumngrid

Recent progress of x-ray scattering techniques enables to measure structure factors of liquid water with high accuracy down to 254.1~K~\cite{skinner2014structure,skinner2013benchmark}, which makes a detailed structural analysis possible even for real water, as for silica.
Here we analyse the O-O partial structure factors of real water as well as TIP4P/2005, TIP5P, and ST2 waters by using four peak functions for fitting (see Fig.~\ref{fig:sqooT}). Indeed, we find that the apparent ``first diffraction peak'' in the O-O partial structure factors of real water as well as model waters can be nicely described by the sum of a Lorentzian (L1) and a Guassian (G1) over a wide temperature range (Figs.~\ref{fig:sqooT} and~\ref{fig:sqoofit}). We call this fitting scheme `scheme II' (see Methods).   In particular, the Lorentzian and Gaussian functions have peaks at $k_\mathrm{T1}=kr_{\mathrm{OO}}/2\pi \sim 3/4$ and $k_\mathrm{D1}=kr_{\mathrm{OO}}/2\pi \sim 1$, corresponding to LFTS and DNLS respectively, in agreement with the above-mentioned silica case and the Debye scattering function shown in Fig.~\ref{fig:sqooTetraZeta}. The integrated intensity of the Lorentzian peak follows the prediction of the two-state model and agrees well with the fraction of LFTS, $s$,  determined independently by $\zeta$ and $N_\mathrm{fs}$. Here we note that the Lorentzian and Gaussian shapes reflect the different nature of the two structural motifs: LFTS has rather unique local tetrahedral order, whereas DNLS intrinsically has high structural fluctuations.
The presence of the bimodality in the experimental structure factor of liquid water (Fig.~\ref{fig:sqooT}), as well as in $\zeta$~\cite{Russo2014,shi2018Microscopic,shi2018impact,shi2018common} and $N_\mathrm{fs}$, together with their inter-consistency, unambiguously show the existence of the two types of local structures (LFTS and DNLS) in liquid water and thus support the two-state description of liquid water.


\begin{table}[t!]
	\caption{Two-state parameters for real water and model waters.}
	\begin{center}
		\begin{tabular}{|c|c|c|c|c|}
			\hline 
			& Real water & TIP4P/2005 & TIP5P & ST2 \\
			\hline
			$\Delta E$ (K)  & -1929.0 & -1802.0 & -3355.9 & -4612.5 \tabularnewline
			\hline 
			$\Delta \sigma$ & -8.2845 & -7.5779 & -13.134 & -16.106 \tabularnewline
			\hline 
			$T_{s=1/2}$     & 232.85  & 237.80 & 255.51 & 286.39 \tabularnewline
			\hline 
		\end{tabular}
	\end{center}
	\label{Tablepara}
\end{table}

Unlike at low $T$, we find that at high $T$ only one Gaussian function is enough to properly describe the apparent ``first diffraction peak'' in the experimental and simulated O-O structure factors. We call this fitting scheme `Scheme I'.  One might think that Scheme I might work even at any temperatures, which is expected for continuum models. Thus, to rationalise the relevance of Scheme II at low $T$, or to confirm the bimodality of the apparently first diffraction peak in an unambiguous manner, we show in Fig.~\ref{fig:fitError} the difference in the mean squared residual, which measures the deviation of the fit from the data, between Schemes I and II as a function of the scaled temperature $T/T_{s=1/2}$. We can clearly see a tendency common to both real water and simulated model waters: 
At temperatures above 1.1$T_{s=1/2}$ a single Gaussian (Scheme I) can describe the apparent ``first diffraction peak'' in the structure factor equally well as a Gaussian plus a Lorentzian function (Scheme II). Below 1.1$T_{s=1/2}$, on the other hand, Scheme I starts to seriously fail in describing the data, reflecting the rapid  growth of the fraction of LFTS below that temperature. The failure of Scheme I at low temperatures not only supports the emergence of the bimodality in the apparent ``first diffraction peak'' there, but also explain why the two-state feature can hardly be observed in liquid water at ambient condition~\cite{smith2005unified,clark2010small,niskanen2019compatibility}.

\begin{center}
	\begin{figure}[b!]
		\includegraphics[width=14cm]{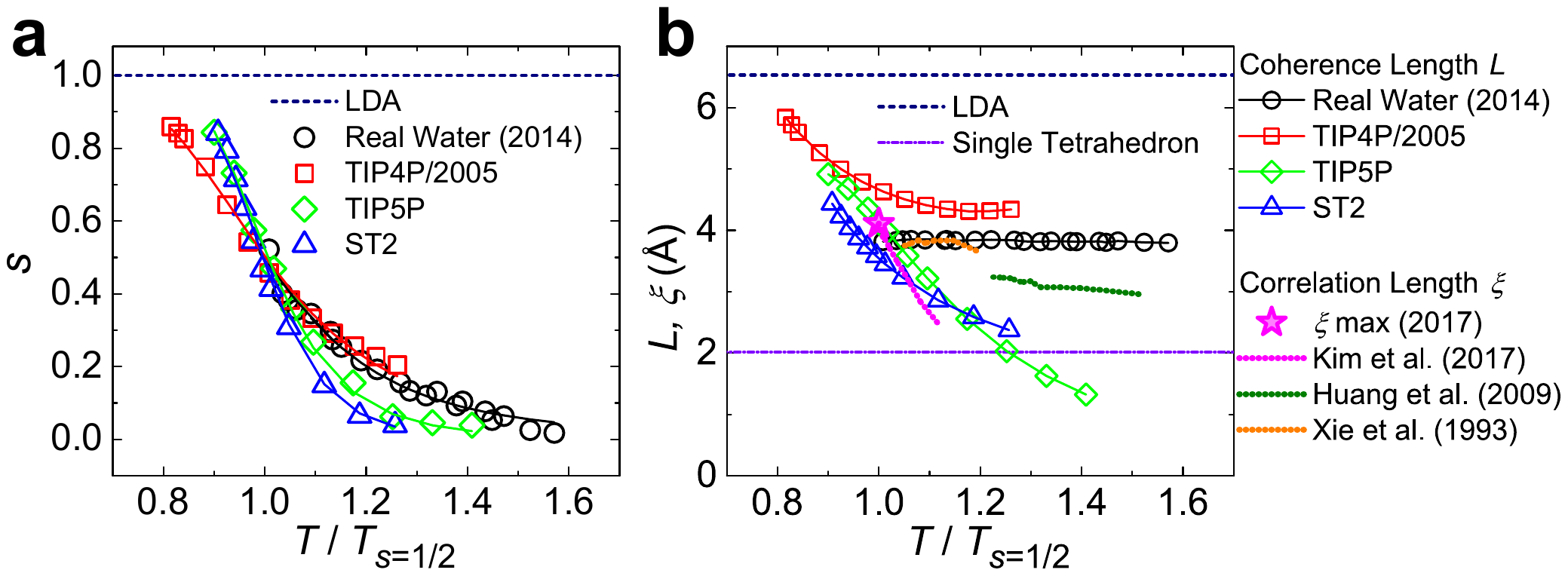}
		\vspace{-0mm}
		\caption{Degree and range of local tetrahedral ordering in liquid water at ambient pressure. {\bf a}, The integrated intensity of FSDP ($k_\mathrm{T1}$ peak in Fig.~\ref{fig:sqooT}b) as a measure of the degree of local tetrahedral ordering, or the fraction $s$ of LFTS, which monotonically increases upon cooling. The horizantal dash line indicates the upper limit of tetrahedral ordering in liquid water. {\bf b}, The coherence length of FSDP as a measure of the range of local tetrahedral ordering, which monotonically increases with decreasing temperature. The high and low temperature limits of the coherence length from a single tetrahedron and LDA ice~\cite{mariedahl2018} are shown by violet dash dot line and navy dot line, respectively (see Fig.~\ref{fig:sqooTetraLDA}). The correlation lengths determined by the Ornstein-Zernike analysis of small-angle X-ray scattering data (typically $k<0.5$~\AA$^{-1}$) by different groups~\cite{xie1993noncritical,huang2009inhomogeneous,kim2017maxima} are shown by dotted lines. The maximum correlation length of $\xi\simeq 4.1$~\AA\,\ at $T_{s=1/2} = 229.2$~K estimated from recent small-angle X-ray scattering measurements of liquid water droplets~\cite{kim2017maxima} is shown by the magenta star symbol.
		}
		\label{fig:sl}
	\end{figure}
\end{center}

Moreover, our two-state description (Scheme II) of the structure factor provides a direct experimental access to the degree and range of local structural ordering in real water. 
The fraction $s$ of LFTS, which is propotional to the integrated intensity of FSDP at $k_\mathrm{T1}$, increases rapidly towards the LDL limit ($s\simeq 1$) upon cooling, as shown in Fig.~\ref{fig:sl}a. The increase is faster for TIP5P and ST2 water than for TIP4P/2005 and the real water, indicating the ``over-structured" tendency in the former two models. In the two-state language, TIP5P and ST2 models overestimate the energy gain and entropy loss upon the formation of LFTS, as shown by the two-state-model parameters in Table~\ref{Tablepara}.

Figure~\ref{fig:sl}b shows the increase of the coherence length estimated from the width of FSDP (Eq.~\ref{eq:L}) upon cooling.  Below $T_{s=1/2}$, the coherence lengths estimated from the experimental and simulated structure factors increase and converge towards the $s \rightarrow 1$ limit upon cooling. Above $T_{s=1/2}$, on the other hand, the fraction of LFTS, i.e. the integrated intensity of FSDP, is rather small and thus the data suffer from large uncertainty. In any case, the coherence length is very short, bounded between $\sim2$~\AA\,\ of a single tetrahedron and $\sim6.5$~\AA\,\ of LDA ice (see Fig.~\ref{fig:sqooTetraLDA} for the detail), in agreement with the previous measurements of structural correlation length~\cite{xie1993noncritical,huang2009inhomogeneous,kim2017maxima} in real water and dynamic correlation length in TIP5P water~\cite{shi2018origin,shi2018common} [Feature (1) in the introduction].

\onecolumngrid
\twocolumngrid

We show the first clear experimentally accessible evidence in the structure factor for the dynamical coexistence of the two types of structural motifs, LFTS and DNLS, supporting the two-state description of liquid water. We reveal that liquid water exhibits the so-called FSDP in the structure factor as other tetrahedral liquids do. The FSDP provides crucial information on the fraction of LFTS (degree of structural ordering, i.e., the order parameter of the two-state model (Eq.~\ref{eq:s})) and its coherence length (range of structural ordering). We hope that these findings will contribute to the convergence of long-standing debates on the structure of water.

\vspace{0.3cm}
\noindent
{\bf Acknowledgements}
\noindent
The authors are grateful to R. Evans for his valuable suggestion on the analysis regarding the criticality.  
This study was partly supported by Scientific Research (A) and Specially Promoted Research (KAKENHI Grants No. JP18H03675 and No. JP25000002 respectively) from the Japan Society for the Promotion of Science (JSPS) and the Mitsubishi Foundation.

\bibliographystyle{naturemag_noURL} 
\bibliography{fsdpWater}

\setcounter{figure}{0}
\renewcommand{\thefigure}{S\arabic{figure}}
\setcounter{table}{0}
\renewcommand{\thetable}{S\arabic{table}}
\setcounter{equation}{0}
\renewcommand{\theequation}{S\arabic{equation}}

\section*{Methods}

\subsection*{Simulation of water}
Classical molecular dynamics simulations were performed in a periodic cubic box containing 1000 TIP4P/2005~\cite{abascal2005general} water molecules by using the Gromacs package~\cite{Hess2008} with a time step of 2~fs. Intermolecular van der Waals forces and Coulomb interactions in real space were truncated at 9~\AA, and long-range Coulomb interactions were treated by the particle-mesh Ewald method. Long-range dispersion corrections for energy and pressure were applied. All simulations were performed in \textit{NPT} ensemble with temperature and pressure kept constant by Nos{\'e}-Hoover thermostat and Parrinello-Rahman barostat, respectively. All the bonds are constrained by using the LINCS algorithm. Long-time simulations (typically longer than 100 times molecular reorientation time) were performed after equilibration in a wide temperature range from 194 to 300 K at 1 bar, and in a wide pressure range from 1 to 10000 bar at 250 K. At 194, 197 and 200 K, two independent trajectories were generated to enhance the statistics. The simulation times for production runs are summarized in Tables~\ref{TabletimeT} and~\ref{TabletimeP}. The simulation details of TIP5P and ST2 model can be found in Refs.~\cite{shi2018origin,shi2018common}. Ice nucleation has not been observed at any temperature and pressure studied in this work for all the water models.

\begin{table}[h!]
	\caption{Simulation times $t$ used for production runs for TIP4P/2005 water at 1 bar. The bold multiplers indicate the number of independent runs.}
	\begin{center}
		\begin{tabular}{|c|c|c|c|c|c|c|c|}
			\hline 
			$T$ (K)& 194 & 197 & 200 & 210 & 220 & 230 & 240\\
			\hline
			$t$ (ns) & $\mathbf{2} \times 38000$ & $\mathbf{2} \times 41000$ & $\mathbf{2} \times 23000$ & 600 & 40 & 10 & 5 \\
			\hline 
			\hline 
			$T$ (K)& 250 & 260 & 270 & 280 & 290 & 300 & \\
			\hline
			$t$ (ns) & 3 & 2 & 1.2 & 1.2 & 1.0 & 1.0 & 
			\tabularnewline
			\hline 
		\end{tabular}
	\end{center}
	\label{TabletimeT}
\end{table}

\begin{table}[h!]
	\caption{Simulation times $t$ used for production runs for TIP4P/2005 water at 250 K.}
	\begin{center}
		\begin{tabular}{|c|c|c|c|c|c|c|c|}
			\hline 
			$P$ (bar)& 1 & 1000 & 1800 & 4000 & 6000 & 8000 & 10000\\
			\hline
			$t$ (ns) & 3 & 2 & 2.6 & 20 & 40 & 80 & 100 \\
			\hline 
		\end{tabular}
	\end{center}
	\label{TabletimeP}
\end{table}

\subsection*{Simulation of silica}
A two-component system of 3456 BKS~\cite{Beest1990,Saika2000} silica (3456 silicon and 6912 oxygen ions) in a periodic cubic box was simulated with a time step of 0.5~fs by using the LAMMPS package~\cite{Plimpton1995}. Short range interactions were truncated at 5.5~\AA, and long-range electrostatic interactions were treated by using the  particle-particle particle-mesh method. All simulations were performed in \textit{NPT} ensemble with temperature and pressure controlled by Nos{\'e}-Hoover thermostat and barostat, respectively. Production runs were carried out at ambient pressure for 200 ps each at 5500 K and 6000 K, after equilibration runs of the same length.

\subsection*{Simulation of metallic glass Cu$_{64}$Zr$_{36}$}
A binary metallic glass Cu$_{64}$Zr$_{36}$, containing 6400 copper and 3600 zirconium atoms was simulated by using the embedded atom method potential~\cite{mendelev2009}. Molecular dynamics simulations were carried out with a time step of 1.0 fs by using the LAMMPS package~\cite{Plimpton1995}. Periodic boundary condition was applied to all directions of the cubic box. Nos{\'e}-Hoover thermostat and barostat were employed to keep the temperature at 1800 K and pressure at 1 bar, respectively. A production run of 400~ps was performed at 1~bar, 1800~K after a 300~ps equilibration run.

\subsection*{Simulation of Lennard-Jones (LJ) liquid}
A one-component LJ system of 6912 particle interacting via the standard 12-6 LJ potential was simulated with a time step of 0.005 by using the LAMMPS package~\cite{Plimpton1995}. The interaction was truncated and force-shifted at $2.5\sigma$, so that both the potential and force smoothly go to zero at $2.5\sigma$. \textit{NVT} and \textit{NPT} simulations were performed at $\rho=0.7$, $t=0.75$ and $p=0.05$, $t=0.75$ (in reduced unit) for 4000000 steps, respectively.

\subsection*{Characterization of local density}
In this work the local density is characterized by the coordination number that measures the number of neighboring water molecules in the first coordination shell of a center molecule. The first coordination shell is defined as a sphere with a radius corresponding to the position of first minimum in the oxygen-oxygen radial distribution function, which is typically $\sim3.5$~\AA, in comparable with the coherence length for liquid water (Fig. 4b).

Other approaches such as Voronoi tessellation~\cite{duboue2015characterization} and density in grids~\cite{soper2010recent,english2011density} have also been used to measure the local density of liquid water. The former estimates local density by using the Voronoi volume of each molecule, whereas the latter calculates the number of molecules in a small cubic box with different sizes (typically $>9$~\AA). Both of these two methods report unimodal density distributions, which have often been taken as direct evidence against the two-state model. However, we argue that neither of the two methods is a proper measure of the local density. For the Voronoi tessellation method, it has been shown that its application to tetrahedral materials such as amorphous silicon suffers from a serious deficiency because of the low coordination number~\cite{tsumuraya1993statistics}. For the grid method, on the other hand, because of the small coherence length of the structural motifs (Fig.~4b), a box of $9 \times 9 \times 9$~\AA\,\ is too large to detect the local density fluctuation associated with them: For a box significantly larger than the size of the local structural motifs, the density distribution is expected to be unimodal and Gaussian, as shown in Fig.~\ref{fig:fdensity}.

\subsection*{Fitting formula for the coordination number distribution}
\begin{figure}[h!]
	\begin{center}
		\includegraphics[width=12cm]{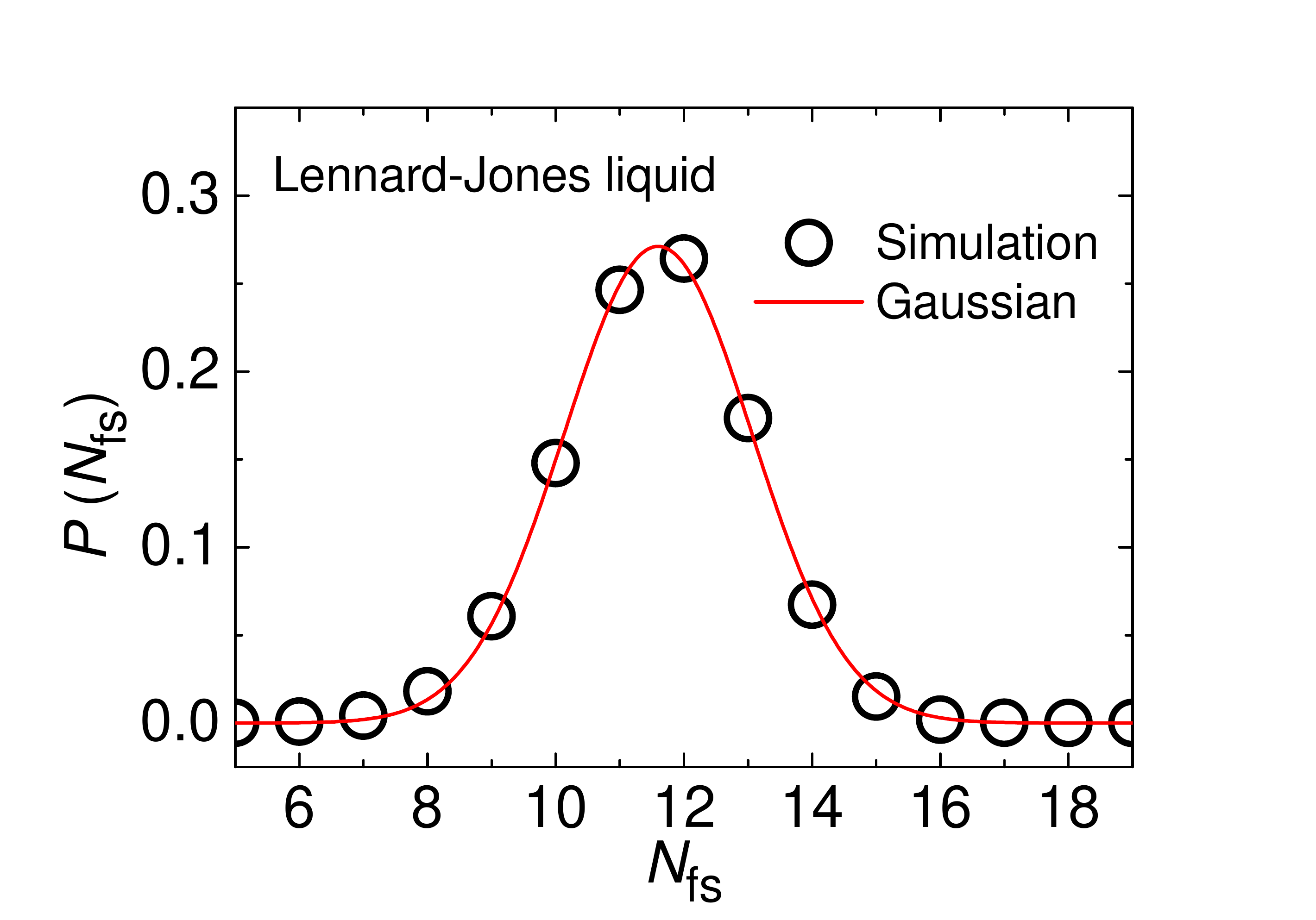}
	\end{center}
	\caption{An example of the typical distribution of coordination number $N_\mathrm{fs}$, $P\left(N_\mathrm{fs}\right)$, for a simple liquid. Here we show $P\left(N_\mathrm{fs}\right)$ of Lennard-Jones liquid at $\rho = 0.7$ and $T=0.75$. $P\left(N_\mathrm{fs}\right)$ can be well described by a Gaussian distribution as indicated by the red curve.}
	\label{fig:fnfs-lj}
\end{figure}

\begin{figure}[h!]
	\begin{center}
		\includegraphics[width=16cm]{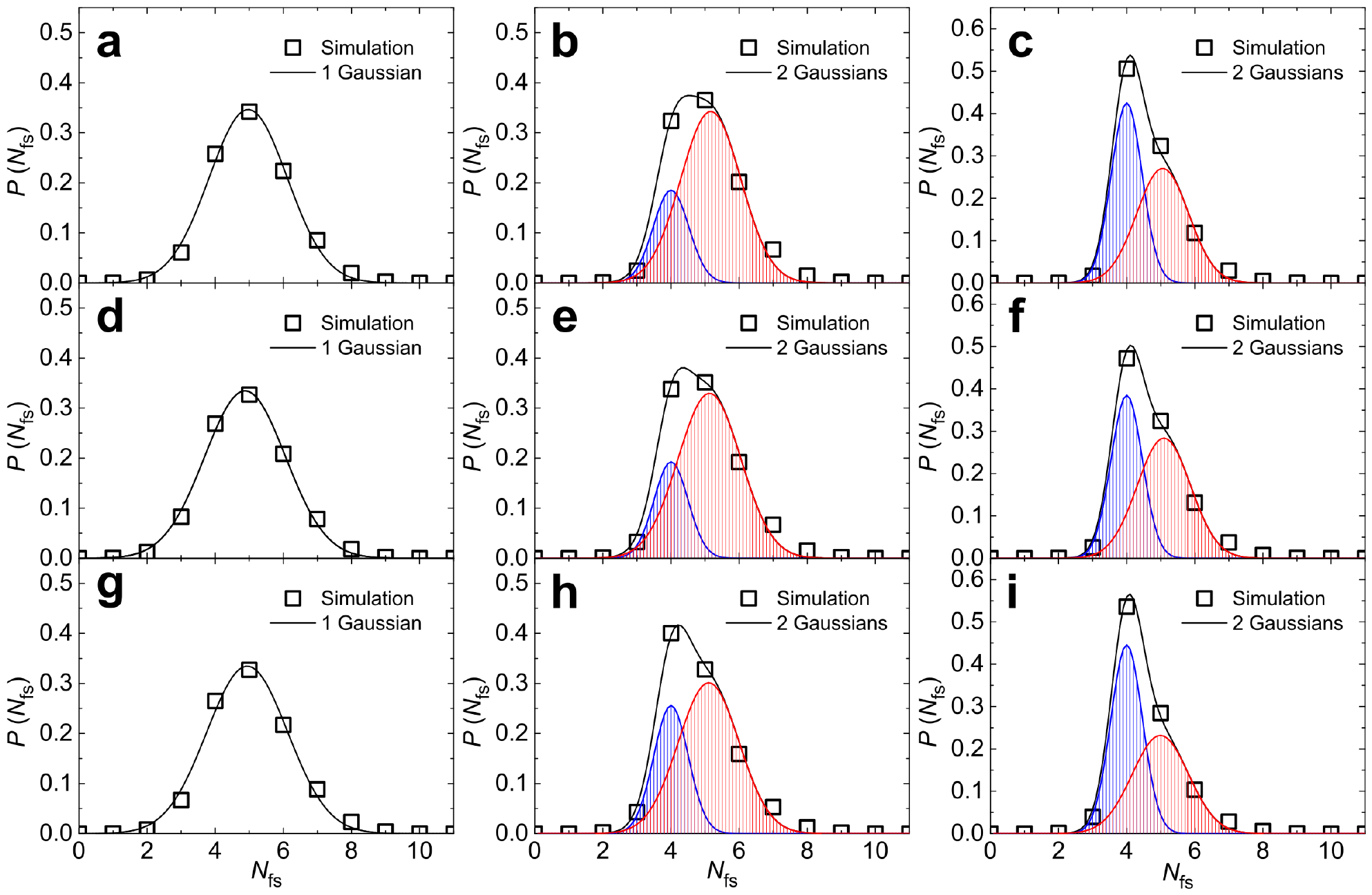}
	\end{center}
	\caption{The distribution of coordination number $N_\mathrm{fs}$, $P\left(N_\mathrm{fs}\right)$, for simulated water. Here we show $P\left(N_\mathrm{fs}\right)$ for TIP4P/2005 water at 380 K ({\bf a}), 280 K ({\bf b}), 240 K ({\bf c}), for TIP5P water at 360 K ({\bf d}), 280 K ({\bf e}), 260 K ({\bf f}), and for ST2 water at 360 K ({\bf g}), 300 K ({\bf h}), 285 K ({\bf i}), at ambient pressure. At high temperatures, $P\left(N_\mathrm{fs}\right)$ shows a Gaussian distribution as simple Lennard-Jones liquid does (Fig.~\ref{fig:fnfs-lj}), whereas at low temperatures it changes to bimodal distributions, which can be properly described by the sum of two Gaussian functions (with blue and red shades). The fraction of the two Gaussian components agrees well with the fraction independently determined by $\zeta$ distribution and the prediction of the theoretical two-state model (Eq.~\ref{eq:s}). One component (with blue shade) corresponds to LFTS, in which the central water typically has $\sim 4$ H-bonded nearest neighbors, whereas the other (with red shade) corresponds to DNLS, in which the central water has $\sim 5$ nearest neighbors (three of them being H-bonded typically) on average at ambient pressure.}
	\label{fig:fnfsT}
\end{figure}

\begin{figure}[h!]
	\begin{center}
		\includegraphics[width=16cm]{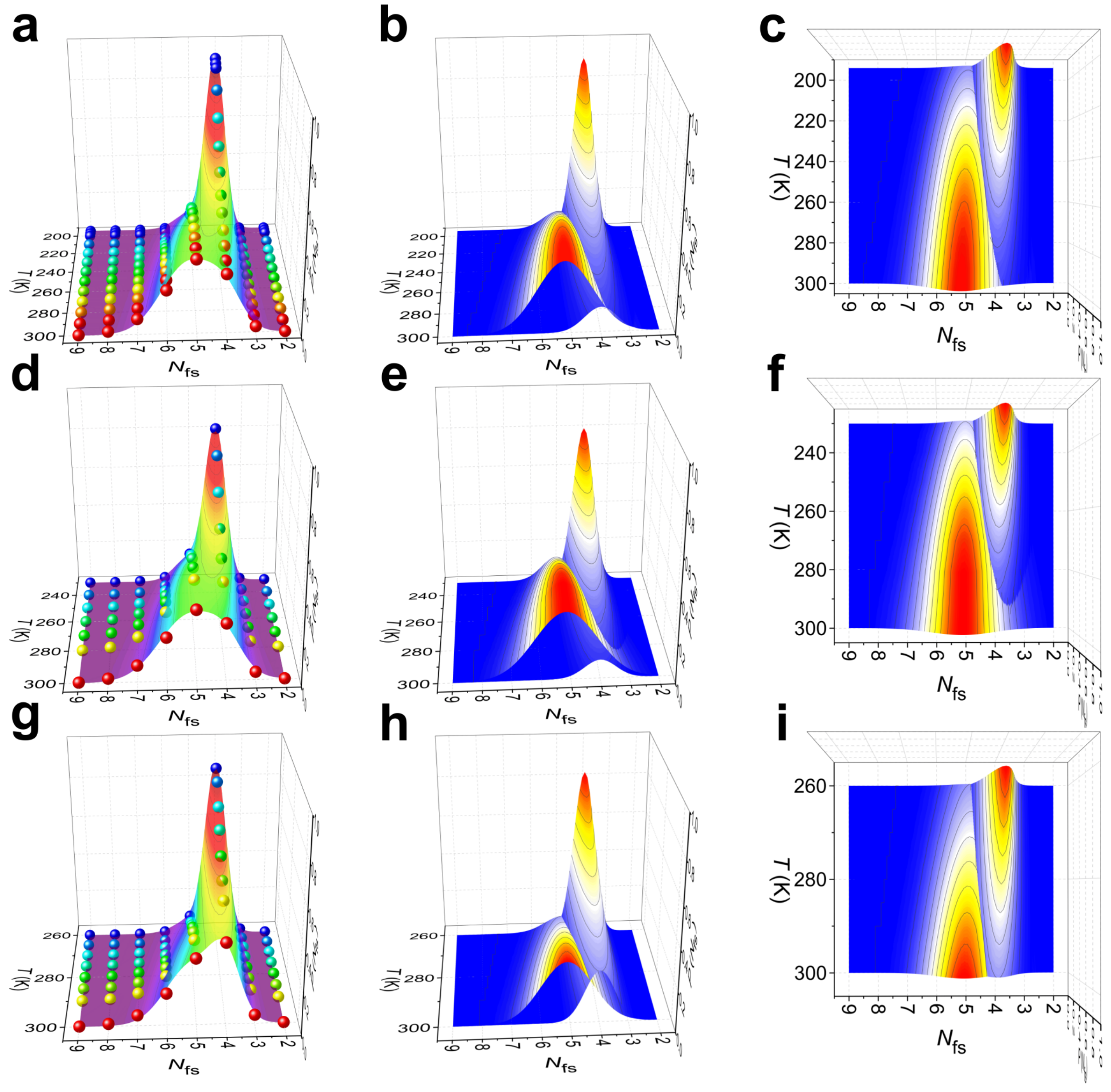}
	\end{center}
	\caption{$T$-dependence of the distribution of coordination number $N_\mathrm{fs}$, $P\left(N_\mathrm{fs}, T\right)$, for three water models. {\bf a}, {\bf d}, {\bf g}, $P \left(N_\mathrm{fs}, T\right)$ for simulated liquid TIP4P/2005 ({\bf a}), TIP5P ({\bf d}) and ST2 ({\bf g}) water at ambient pressure.  The colored surfaces in {\bf a}, {\bf d}, {\bf g} are the fits to two Gaussian functions by Eqs.~\ref{eq:pnfs} - \ref{eq:SDNLS}. The two Gaussian components of $P \left(N_\mathrm{fs}, T\right)$ are shown in side ({\bf b}, {\bf e}, {\bf h}) and top ({\bf c}, {\bf f}, {\bf i}) view for TIP4P/2005 ({\bf b}, {\bf c}), TIP5P ({\bf e}, {\bf f}) and ST2 ({\bf h}, {\bf i}) water.}
	\label{fig:fnfs2Gau-small}
\end{figure}

\begin{figure}[h!]
	\begin{center}
		\includegraphics[width=14cm]{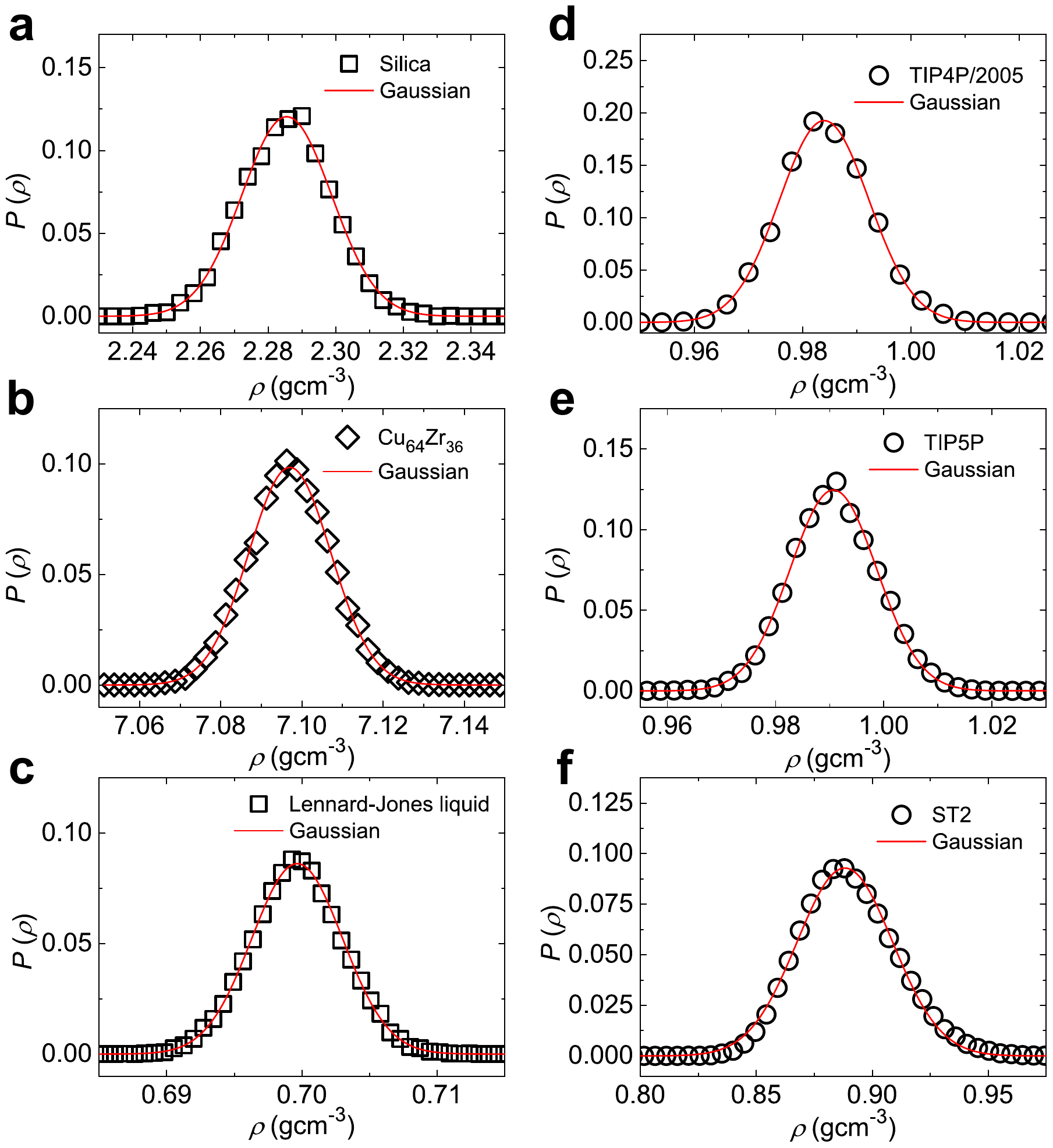}
	\end{center}
	\caption{Macroscopic density distribution for single- and two-component systems. Macroscopic density distribution $P(\rho)$ for two-component BKS silica at $T=6000$~K, $P=1$~bar ({\bf a}), two-component metallic glass Cu$_{64}$Zr$_{36}$ at $T=1800$~K, $P=1$~bar ({\bf b}), single-component Lennard-Jones liquid at $T=0.75$, $P=0.05$ ({\bf c}) and single-component TIP4P/2005 water at  $T=240$~K, $P=1$~bar ({\bf d}), TIP5P water at $T=260$~K, $P=1$~bar ({\bf e}) and ST2 water at $T=285$~K, $P=1$~bar ({\bf f}). The macroscopic density distributions in panels {\bf d}, {\bf e} and {\bf f} were measured at the same conditions as in panels {\bf c}, {\bf f} and {\bf i} in Fig.~\ref{fig:fnfsT} for TIP4P/2005, TIP5P and ST2 water, respectively. The macroscopic density distribution always remains unimodal and Gaussian for both single- and two-component liquids, under thermal fluctuations, as it should be.}
	\label{fig:fdensity}
\end{figure}

\begin{figure}[h!]
	\begin{center}
		\includegraphics[width=14cm]{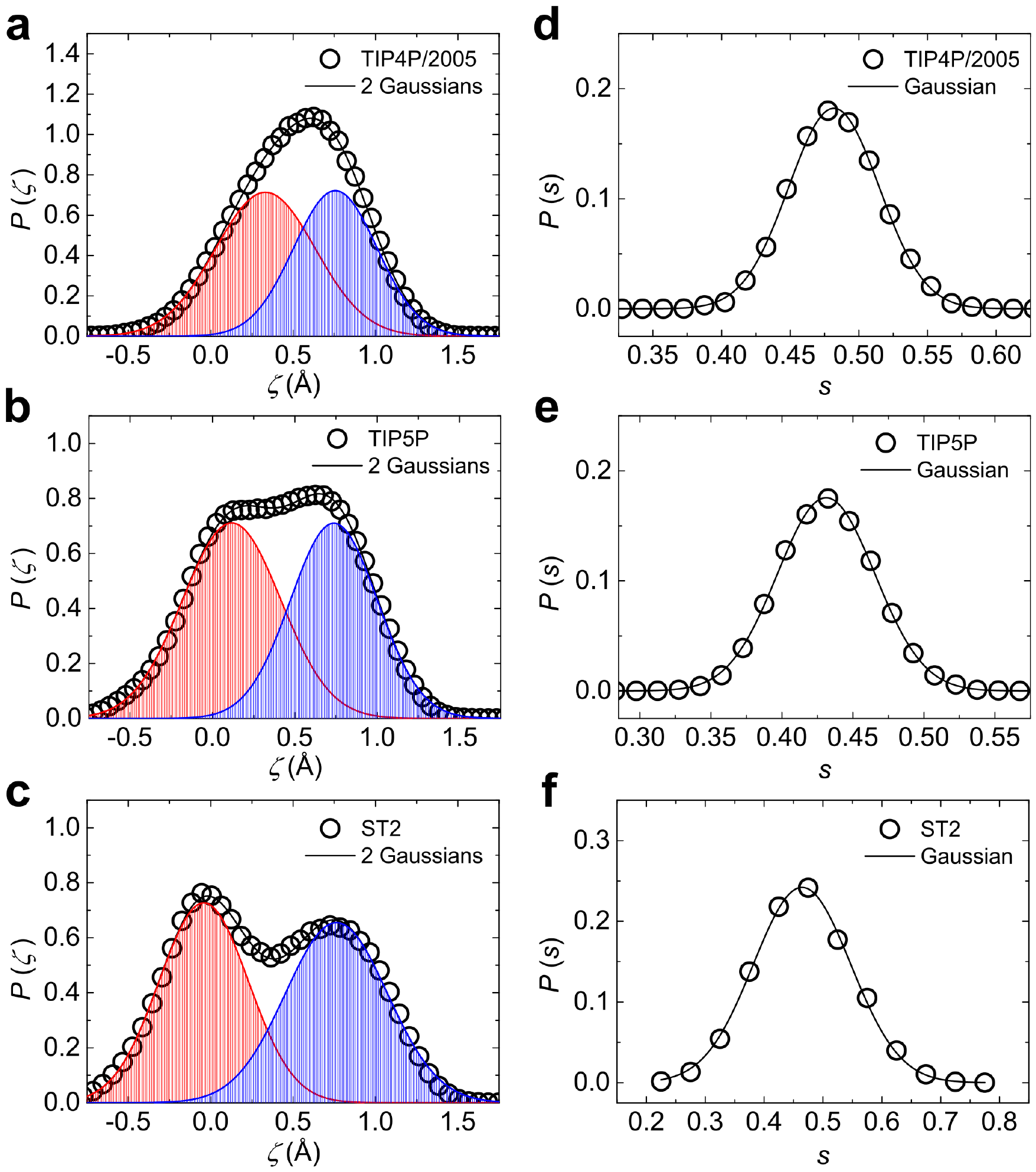}
	\end{center}
	\caption{Distribution of local structural descriptor $\zeta$ and macroscopic order parameter $s$. {\bf a}-{\bf c}, Distribution of local structural descriptor $\zeta$, $P(\zeta)$, for TIP4P/2005 ({\bf a}), TIP5P ({\bf b}) and ST2 water ({\bf c}) at ambient pressure. {\bf d}-{\bf f}, Distribution of macroscopic structural order parameter $s$, $P(s)$, for TIP4P/2005 ({\bf d}), TIP5P ({\bf e}) and ST2 water ({\bf f}) at ambient pressure. All the distributions are shown at $T\approx T_{s=1/2}$ (240~K for TIP4P/2005 water, 260~K for TIP5P water and 285~K for ST2). The distribution of the macroscopic order parameter, $P(s)$, always remains unimodal and Gaussian for all the models, whereas the distribution of local order, $P(\zeta)$, is clearly bimodal with two Gaussian components (blue and red shades correspond to LFTS and DNLS, respectively). The macroscopic order parameter $s$ is estimated as the fraction of molecules whose $\zeta > \zeta_c$ at each time frame. A threshold value $\zeta_c$ ($\simeq 0.5$~\AA) is chosen to satisfy $s\approx 0.5$ after time averaging.}
	\label{fig:fs_zeta}
\end{figure}

\begin{figure}[h!]
	\begin{center}
		\includegraphics[width=12cm]{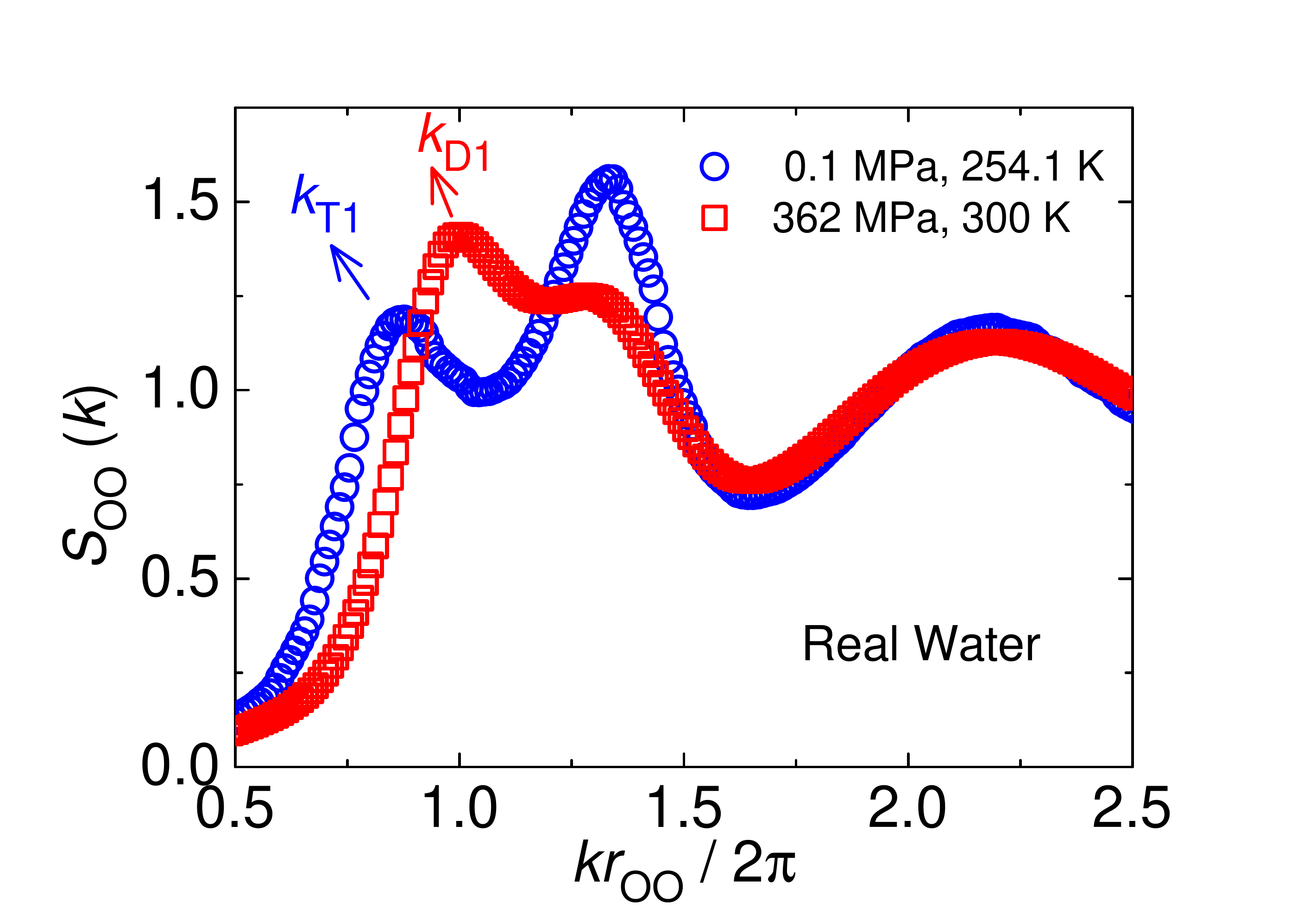}
	\end{center}
	\caption{Partial O-O structure factors of liquid water measured by x-ray scattering experiments at 0.1~MPa, 254.1~K~\cite{skinner2014structure} (blue circle) and 362~MPa, 300~K~\cite{skinner2016structure} (red square). Distinct characteristic peaks $k_\mathrm{T1}$ and $k_\mathrm{D1}$ corresponding to LFTS and DNLS respectively, are clearly located at different wave numbers for these two conditions, whereas the other peaks are located at almost the same positions.}
	\label{fig:sqoo_lowT_highP}
\end{figure}

\begin{figure}[h!]
	\begin{center}
		\includegraphics[width=10cm]{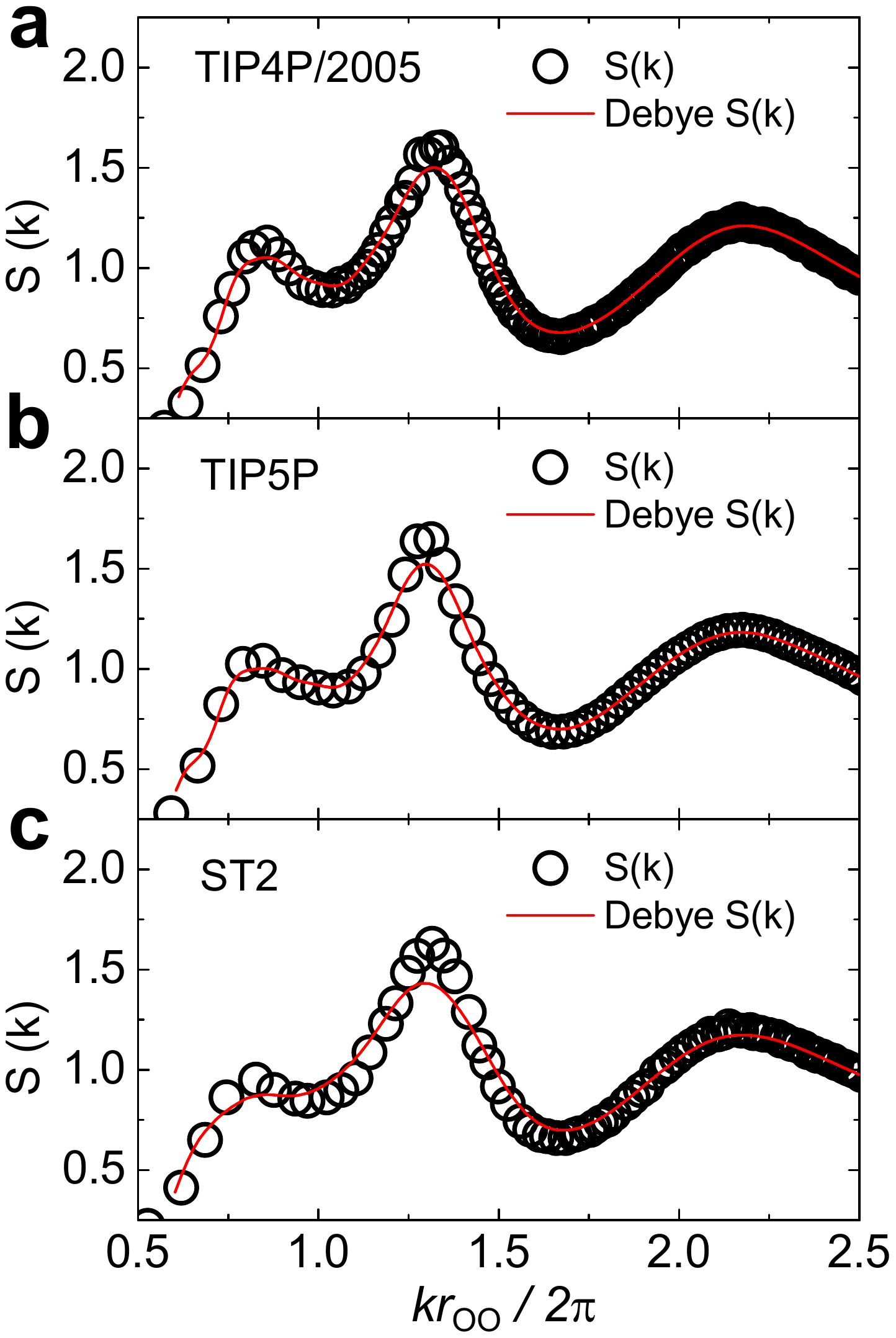}
	\end{center}
	\caption{The validity of the Debye's scattering function analysis. The parital O-O structure factors obtained by the two-body density correlation (Eq.~\ref{eq:skS}, black circle) and Debye scattering equation (Eq.~\ref{eq:skdebye}, red curve) for TIP4P/2005 ({\bf a}), TIP5P ({\bf b}) and ST2 ({\bf c}) water at 240 K, 260 K and 285 K, respectively, at ambient pressure. The structure factors calculated from the two methods agree well with each other. A small deviation at the second peak $kr_\mathrm{OO}/2\pi\approx 1.3$ mainly comes from the effect of the Window function we employed (see Methods).}
	\label{fig:sqdebye}
\end{figure}

\begin{figure}[h!]
	\begin{center}
		\includegraphics[width=14cm]{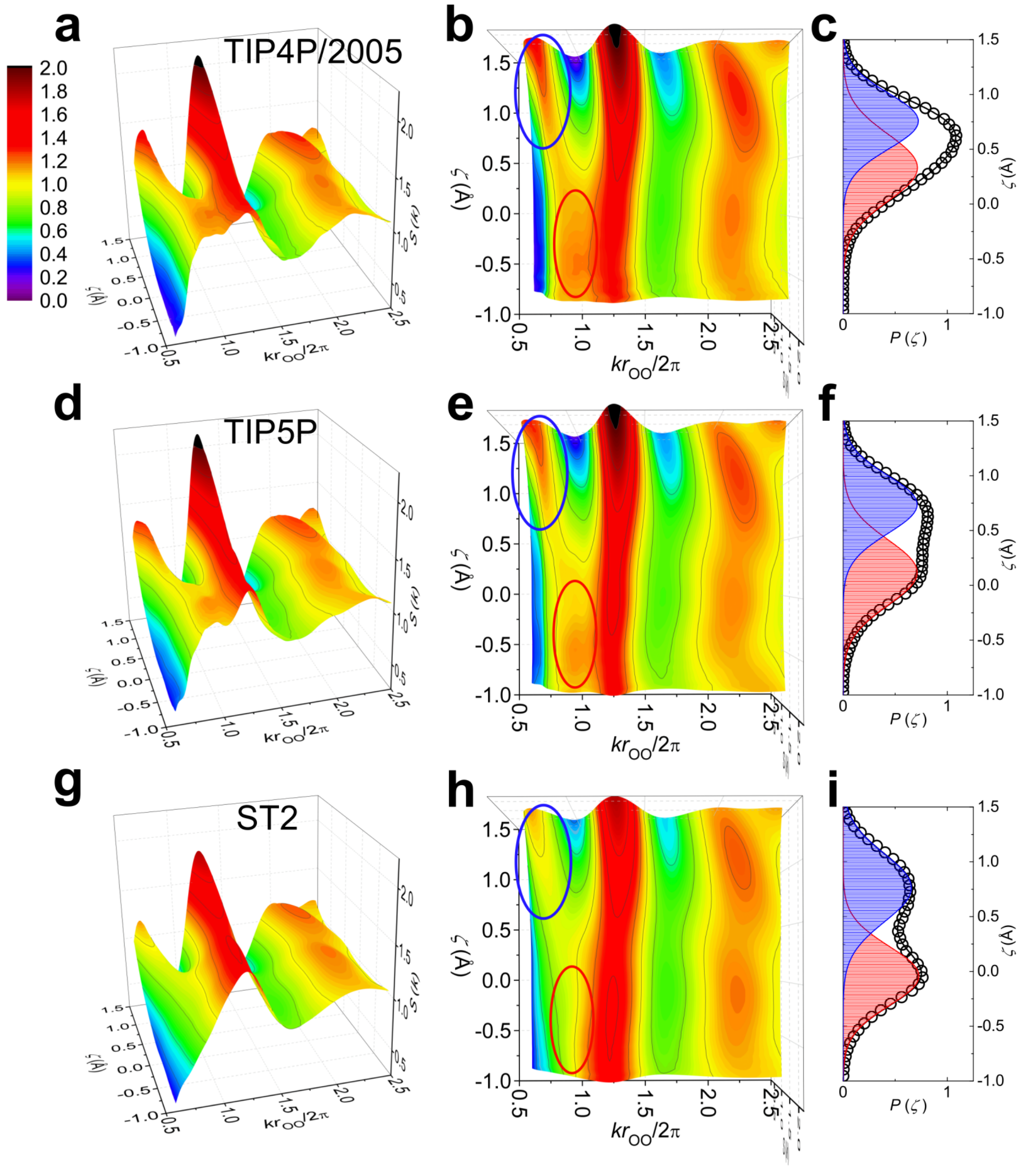}
	\end{center}
	\caption{The $\zeta$ dependent partial O-O structure factor $S(k,\zeta)$ obtained by Debye's scattering equation (Eqs.~\ref{eq:ski} and \ref{eq:skzeta}). Side ({\bf a}, {\bf d}, {\bf g}) and top ({\bf b}, {\bf e}, {\bf h}) views for TIP4P/2005 ({\bf a}, {\bf b}) TIP5P ({\bf d}, {\bf e}) and ST2 ({\bf g}, {\bf h}) water at 240 K, 260 K and 285 K, respectively, at ambient pressure. In ({\bf b}, {\bf e}, {\bf h}) the characteristic peaks, $k_\mathrm{T1}$ and $k_\mathrm{D1}$, are highlighted by blue and red circles, respectively. {\bf c},{\bf f},{\bf i}, The distribution of $\zeta$ shows two Gaussian components, corresponding to LFTS (blue shade) and DNLS (red shade) respectively for TIP4P/2005 ({\bf c}), TIP5P ({\bf f}) and ST2 ({\bf i}) water. The wave numbers are scaled by the nearest neighbor O-O distance $r_\mathrm{OO}$.}
	\label{fig:sqooZeta}
\end{figure}

At high enough temperatures, the distribution of coordination number, $P\left(N_\mathrm{fs}\right)$, of liquid water shows a unimodal Gaussian distribution as simple LJ liquid (see Fig.~\ref{fig:fnfs-lj} and Fig.~\ref{fig:fnfsT}a,d,g). However, $P\left(N_\mathrm{fs}\right)$ of liquid water significantly deviates from a single Gaussian distribution and instead displays a bimodal distribution upon cooling, which strongly suggests the development of two structural motifs in supercooled water (see Fig.~1c and Fig.~\ref{fig:fnfs2Gau-small}).  As a result, we find that $P\left(N_\mathrm{fs}\right)$ can be properly described by the sum of two Gaussian functions, whose integrated intensity corresponds to the fraction of LFTS and DNLS respectively,
\begin{equation}
\begin{split}
P\left(N_\mathrm{fs}\right) & = \frac{s}{\sigma_\mathrm{LFTS} \sqrt{2\pi}} \exp \left[-\frac{\left(N_\mathrm{fs}-N_\mathrm{LFTS}\right)^2}{2\sigma_\mathrm{LFTS}^2}\right] \\ & + \frac{1-s}{\sigma_\mathrm{DNLS} \sqrt{2\pi}} \exp \left[-\frac{\left(N_\mathrm{fs}-N_\mathrm{DNLS}\right)^2}{2\sigma_\mathrm{DNLS}^2}\right]
\label{eq:pnfs}
\end{split}
\end{equation}
In this equation, $s$ is defined by Eq.~\ref{eq:s} and other parameters can be described as follows,
\begin{equation}
N_\mathrm{LFTS} = a_{00} + a_{01}T + a_{02}T^2
\label{eq:NLFTS}
\end{equation}
\begin{equation}
N_\mathrm{DNLS} = a_{10} + a_{11}T + a_{12}T^2
\label{eq:NDNLS}
\end{equation}
\begin{equation}
\sigma_\mathrm{LFTS} = b_{00} + b_{01}T + b_{02}T^2
\label{eq:SLFTS}
\end{equation}
\begin{equation}
\sigma_\mathrm{DNLS} = b_{10} + b_{11}T + b_{12}T^2
\label{eq:SDNLS}
\end{equation}

\subsection*{Calculation of the structure factor}
The structure factor $S(\bm{k})$, defined as the density-density correlation in reciprocal space, can be obtained by 
\begin{equation}
S(\bm{k}) = \frac{1}{N}\langle\rho_{\bm{k}} \rho_{-\bm{k}}\rangle
\label{eq:skS}
\end{equation}
where $\langle \cdots \rangle$ represents the ensemble average, $\bm{k}$ is the wave vector, $N$ is the number of particles and $\rho_{\bm{k}}$ is the Fourier component of the number density $\rho$, which is given by
\begin{equation}
\rho_{\bm{k}} = \stackrel[i=1]{N}{\sum}\exp \left(-i\bm{k}\cdot \bm{r}_{i}\right)
\label{eq:rhok}
\end{equation}
where $\bm{r}_i$ is the coordinates of atom $i$. For an isotropic system, the structure factor is a function of only the magnitude of the wave vector, $k$: $S(k)$.

\subsection*{Debye scattering function}
The structure factor can also be calculated by the Debye scattering function~\cite{debye1915zerstreuung}:
\begin{equation}
S(k) = 1 + \frac{1}{N}  \stackrel[i=1]{N}{\sum}  \stackrel[j\neq i]{N}{\sum} \frac{\sin (kr_{ij})}{kr_{ij}} W(r_{ij})
\label{eq:skdebye}
\end{equation}
where $W(r_{ij}) = \frac{\sin (\pi r_{ij}/r_{c})}{\pi r_{ij}/r_{c}}$ is the window function~\cite{zhang2019improving} and $r_{c}$ is a cutoff distance.

Debye scattering function allows for a local structural characterization by the molecular structure factor: 
\begin{equation}
S_{i}(k) = 1 + \stackrel[j\neq i]{N}{\sum} \frac{\sin (kr_{ij})}{kr_{ij}} W(r_{ij})
\label{eq:ski}
\end{equation}

Then, the correlation between molecular structure factor $S_{i}(k)$ and local structure descriptor $\zeta$ can be evaluated by the $\zeta$-dependent structure factor: 
\begin{equation}
S(k,\zeta) = \frac{\stackrel[i]{N}{\sum} S_{i}(k)\delta(\zeta-\zeta(i))}{\stackrel[i]{N}{\sum} \delta(\zeta-\zeta(i))}
\label{eq:skzeta}
\end{equation}

\subsection*{Fitting formula for the structure factor}

The first three peaks, the first of which is actually a doublet, in the structure factor of liquid water can be well described by two Lorentzian and two Gaussian functions as 
\begin{equation}
\begin{aligned}
S\left(k\right) = \frac{f_\mathrm{T1}}{\pi}\frac{\Gamma_\mathrm{T1}}{\left(k-k_\mathrm{T1}\right)^2 + \Gamma_\mathrm{T1}^2} + \frac{f_\mathrm{D1}}{\sigma_\mathrm{D1} \sqrt{2\pi}} \exp \left[-\frac{\left(k-k_\mathrm{D1}\right)^2}{2\sigma_\mathrm{D1}^2}\right] \\+ \frac{f_\mathrm{T2}}{\pi}\frac{\Gamma_\mathrm{T2}}{\left(k-k_\mathrm{T2}\right)^2 + \Gamma_\mathrm{T2}^2} + \frac{f_\mathrm{T3}}{\sigma_\mathrm{T3} \sqrt{2\pi}} \exp \left[-\frac{\left(k-k_\mathrm{T3}\right)^2}{2\sigma_\mathrm{T3}^2}\right]
\label{eq:sk2}
\end{aligned}
\end{equation}
where the subscripts denote the peaks in the O-O partial structure factor as shown in Fig.~3b. In this equation, all the parameters depend on temperature and pressure, and therefore a large set of parameters are needed to describe the structure factor of liquid water at different thermodynamic conditions (12 parameters for each temperature and pressure).

However, thanks to the weak temperature dependence of the fitting parameters (except for $f_\mathrm{T1}$ and $f_\mathrm{D1}$), we found that they can be well described by a set of polynomial functions up to the second order:
\begin{equation}
k_{x} (T) = \tilde{k}_{x0} + \tilde{k}_{x1} \hat{T} + \tilde{k}_{x2} \hat{T}^2
\label{eq:polyk}
\end{equation}
\begin{equation}
\Gamma_{x} (T) = \tilde{\Gamma}_{x0} + \tilde{\Gamma}_{x1} \hat{T} + \tilde{\Gamma}_{x2} \hat{T}^2
\label{eq:polyg}
\end{equation}
\begin{equation}
\sigma_{x} (T) = \tilde{\sigma}_{x0} + \tilde{\sigma}_{x1} \hat{T} + \tilde{\sigma}_{x2} \hat{T}^2
\label{eq:polys}
\end{equation}
\begin{equation}
f_{x} (T) = \tilde{f}_{x0} + \tilde{f}_{x1} \hat{T} + \tilde{f}_{x2} \hat{T}^2
\label{eq:polyf}
\end{equation}
where the subscript $x$ ($x = \textrm{T1, D1, T2 or T3}$) denotes parameters for each peak and $\hat{T}=T/T_\mathrm{ref}$ with $T_\mathrm{ref} = 373.15$~K. This procedure allows for a simultaneous fitting of a large set of structure factors measured in a wide temperature range, which largely reduces the number of fitting parameters. Moreover, we found in practice that the fitting accuracy will not be affected if we set $\tilde{k}_\mathrm{x2}=0$ (for $x = \textrm{T1, D1, T2 or T3}$), and fix the value of $k_\mathrm{T3}$ and $\tilde{k}_\mathrm{T11}/\tilde{k}_\mathrm{D11}$ properly. Although the $k_\mathrm{T3}$ peak is only partially included in the fitting, its position is read from the data and fixed in the fitting precedure.

On the other hand, the intensities of T1 and D1 peaks vary significantly with temperature and pressure, corresponding to the change in the fractions of the two structural motifs. In our previous study~\cite{shi2019distinct}, we found that the integrated intensity $f_{T1}$ of FSDP is proportional to the fraction $s$ of LFTS in liquid silica as
\begin{equation}
f_\mathrm{T1} (T,P) = a\cdot s
\label{eq:ft1}
\end{equation}
where $a$ is a positive constant. This knowledge is directly applied to liquid water, since they are both characterized by the same two-state features~\cite{shi2019distinct}.

Here $s$ can further be described by the two-state model with negligible cooperativity as~\cite{Tanaka_review,Tanaka2000,russo2018water,shi2018impact,shi2018Microscopic,shi2018origin,shi2018common}: 
\begin{equation}
s(T,P) = \frac{1}{1+\exp \left(\frac{\Delta E - T \Delta \sigma + P \Delta V}{k_\mathrm{B}T}\right)}
\label{eq:s} 
\end{equation}
where $k_\mathrm{B}$ is the Boltzmann constant, $\Delta E$, $\Delta \sigma$ and $\Delta V$ are the energy, entropy, and volume differences between LFTS and DNLS in the two-state model. At ambient pressure, the term $P\Delta V$ is negligibly small. The parameters $\Delta E$ and $\Delta \sigma$ for TIP5P and ST2 waters have already been determined in our previous work~\cite{shi2018common}. For TIP4P/2005 and real water, $\Delta E$ and $\Delta \sigma$ were independently determined by applying the two-state model (Eq.~\ref{eq:s}) to the fraction $s(T)$ of LFTS that can be obtained from the $g_\mathrm{OO}(r)$ by $s(T)=1-g_\mathrm{OO}(r=r_\mathrm{HB})$~\cite{Russo2014}, where $r_\mathrm{HB}=3.5$~\AA, according the Luzar-Chandler definition of H-bond~\cite{Luzar1996hydrogen}. The parameters $\Delta E$ and $\Delta \sigma$ for real water, TIP4P/2005, TIP5P and ST2 waters are summarised in Table~1 in the main text.

Since $D1$ peak is exclusively from DNLS, whose fraction is given by $(1-s)$, we can formulate the temperature and pressure dependence of $f_{D1}$ as
\begin{equation}
f_{D1} (T,P) = b(1-s)=b \left[1- \frac{1}{1+\exp \left( \frac{\Delta E - T \Delta \sigma + P \Delta V}{k_\mathrm{B}T}\right)}\right]
\label{eq:fd1} 
\end{equation}
where $b$ is a positive constant. After all the above considerations, only 25 free fitting parameters are necessary to fit O-O partial structure factors of liquid water at all the temperatures studied in this work.

The Gaussian function represents the scattering peak coming from the interatomic correlation of DNLS (see main text). Simple liquids such as LJ and hard spheres liquids usually have this peak at the wave number corresponding to the neighboring interactomic distance $r$, and thus we constrained its position $k_{0}r_\mathrm{OO}/2\pi$ to be close to 1. The Fourier transform of a Lorentzian function is an exponentially decaying function in real space. The coherence length $\lambda$ of FSDP, which characterizes the range of coherent tetrahedral ordering, can be estimated by
\begin{equation}
\lambda=1/\Gamma_\mathrm{T1}
\label{eq:L}
\end{equation}
where $\Gamma_\mathrm{T1}$ is the half width of FSDP.

\clearpage
\onecolumngrid


\begin{figure}[h!]
	\begin{center}
		\includegraphics[width=12cm]{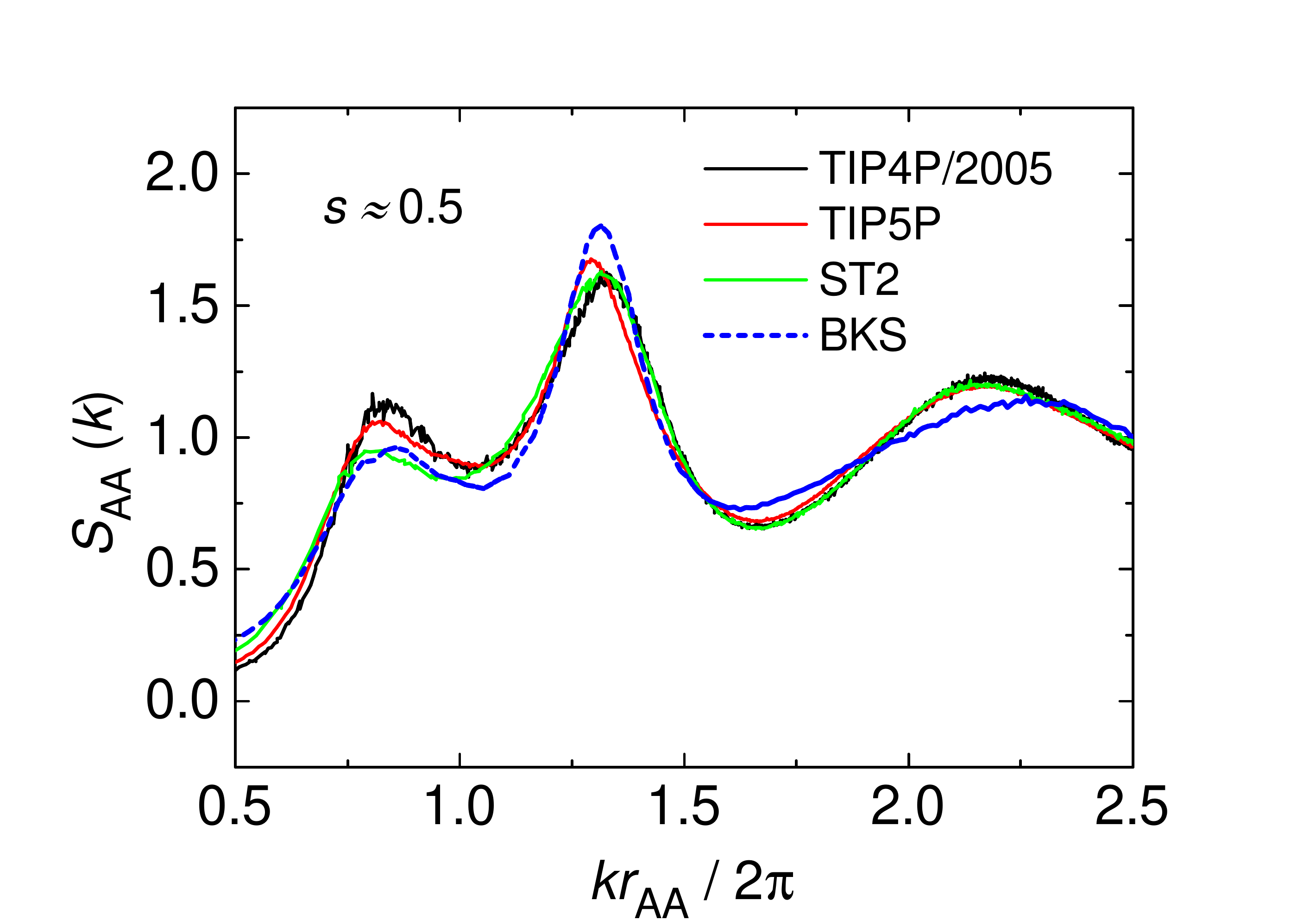}
	\end{center}
	\caption{Partial structure factor $S_\mathrm{AA}(k)$ of simulated water (A=O) and silica (A=Si) at $T\approx T_{s=1/2}$. At ambient pressure, $T_{s=1/2} \approx $ 240, 260, 285 and 5000 K for TIP4P/2005, TIP5P, ST2 water and BKS silica, respectively. Silica shows very similar anomalous behaviors as water does, although no critical point has been found in silica~\cite{lascaris2014search}. Recently strong evidence supporting the existence of two structural motifs---LFTS and DNLS, in liquid silica has been revealed for BKS silica~\cite{shi2018impact,shi2019distinct}. The similarity between the structure factor of liquid water and liquid silica at the same fraction of LFTS further supports the two-state feature in both liquids.}
	\label{fig:sqWaterSilicaTw}
\end{figure}

\begin{figure}[h!]
	\begin{center}
		\includegraphics[width=10cm]{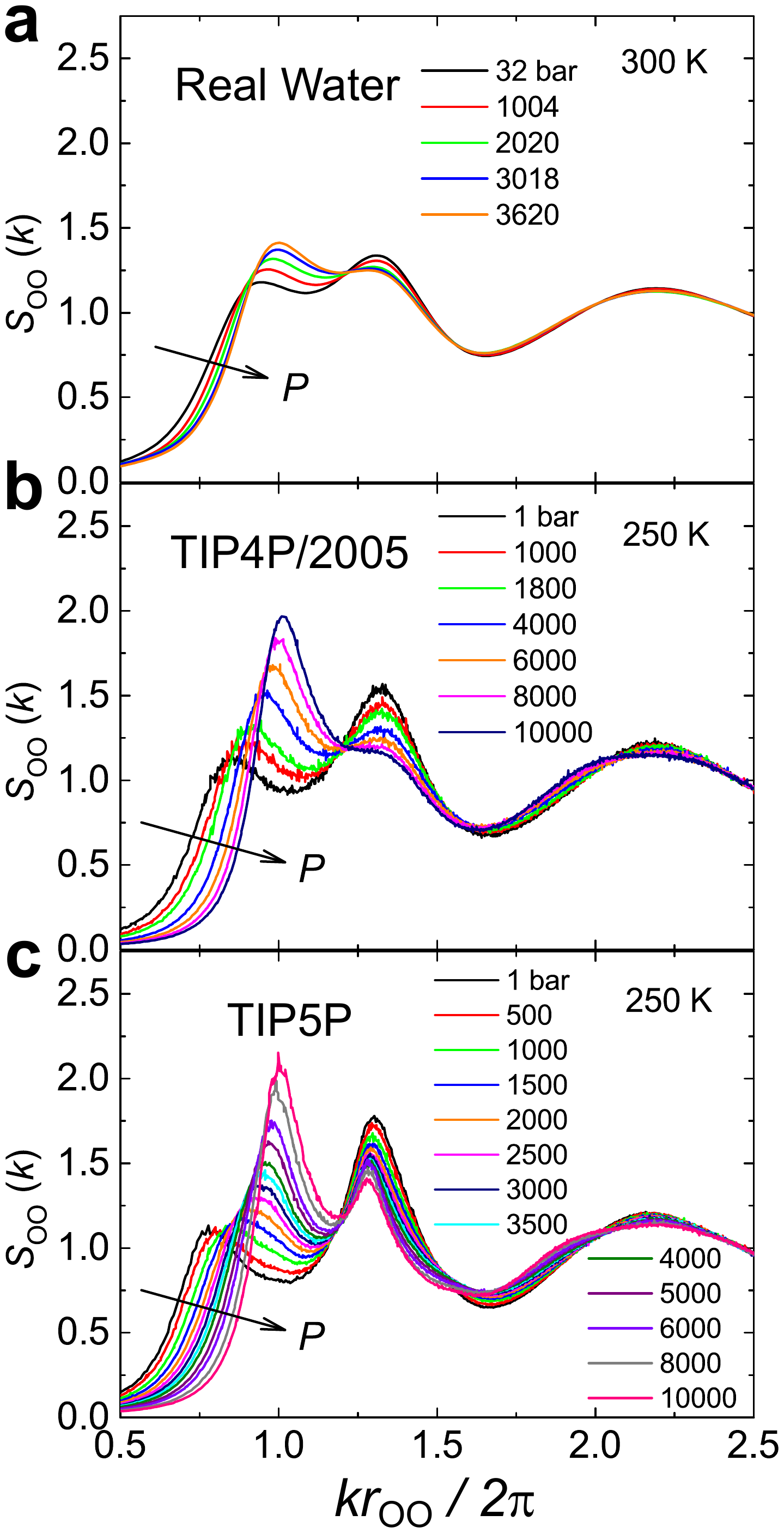}
	\end{center}
	\caption{Experimental O-O partial structure factor of water at high pressure. {\bf a}, X-ray scattering data~\cite{skinner2016structure}. {\bf b}, TIP4P/2005 model. {\bf c}, TIP5P model. The arrow denotes the direction of pressure increase. The wave number is scaled by the nearest neighbor O-O distance $r_\mathrm{OO}$.}
	\label{fig:sqooP}
\end{figure}

\begin{figure}[t!]
	\begin{center}
		\includegraphics[width=16cm]{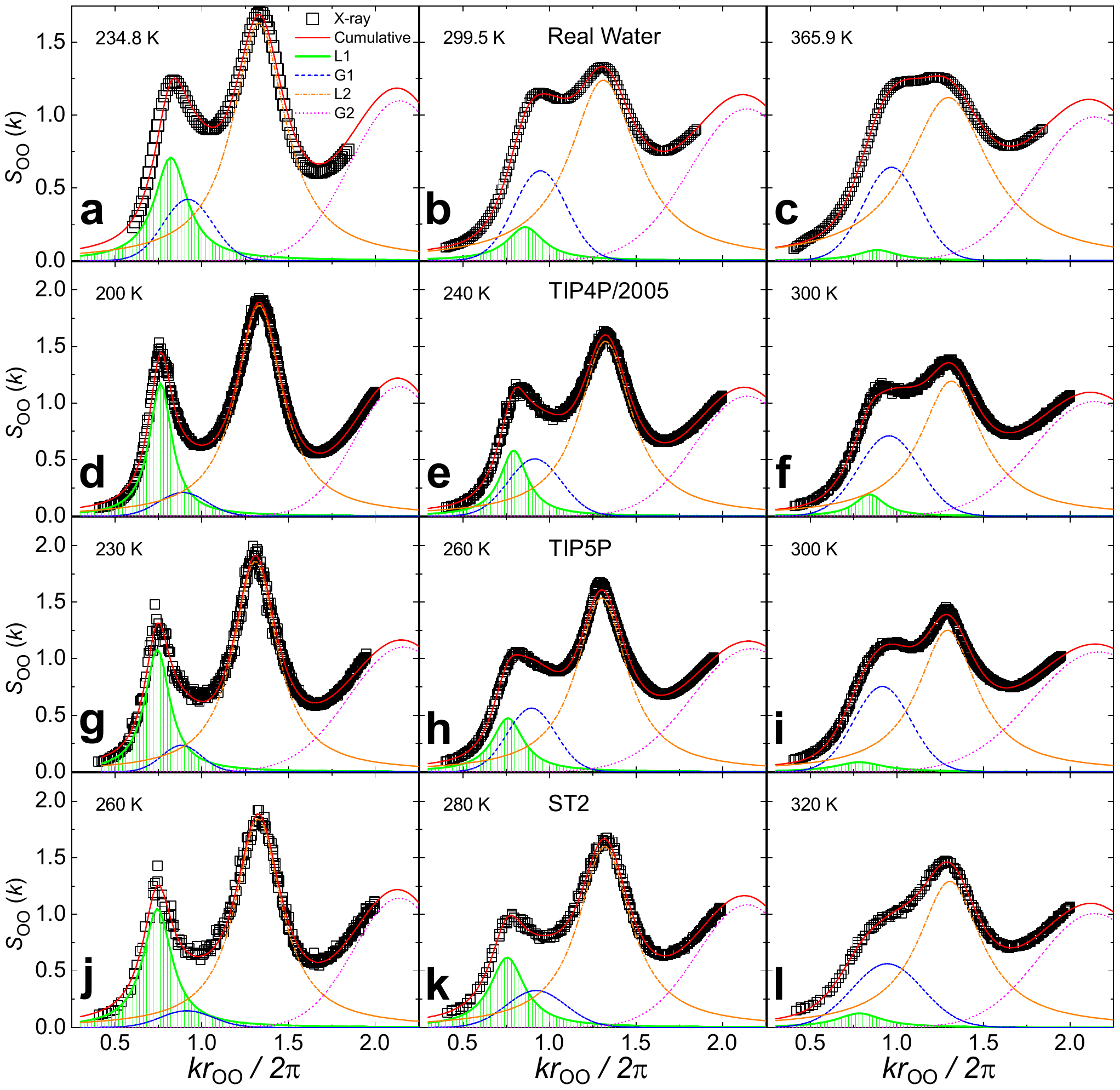}
	\end{center}
	\caption{Decomposition of the O-O partial structure factor of real water and model waters at ambient pressure. {\bf a}-{\bf c}, X-ray scattering data~\cite{skinner2014structure,skinner2013benchmark,pathak2019intermediate}. {\bf d}-{\bf f}, TIP4P/2005 water. {\bf g}-{\bf i}, TIP5P water. {\bf j}-{\bf l}, ST2 water. The O-O partial structure factors are shown by black squares. The red lines are the cumulative fits of two Lorentzian (L1 + L2) and two Gaussian (G1 + G2) functions to the structure factors. The four characteristic peaks obtained from the fits are displayed by green, blue, orange and magenta curves. The FSDP is highlighted by green shading. The wave number is scaled by the nearest neighbor O-O distance $r_\mathrm{OO}$.}
	\label{fig:sqoofit}
\end{figure}

\begin{figure}[h!]
	\begin{center}
		\includegraphics[width=12cm]{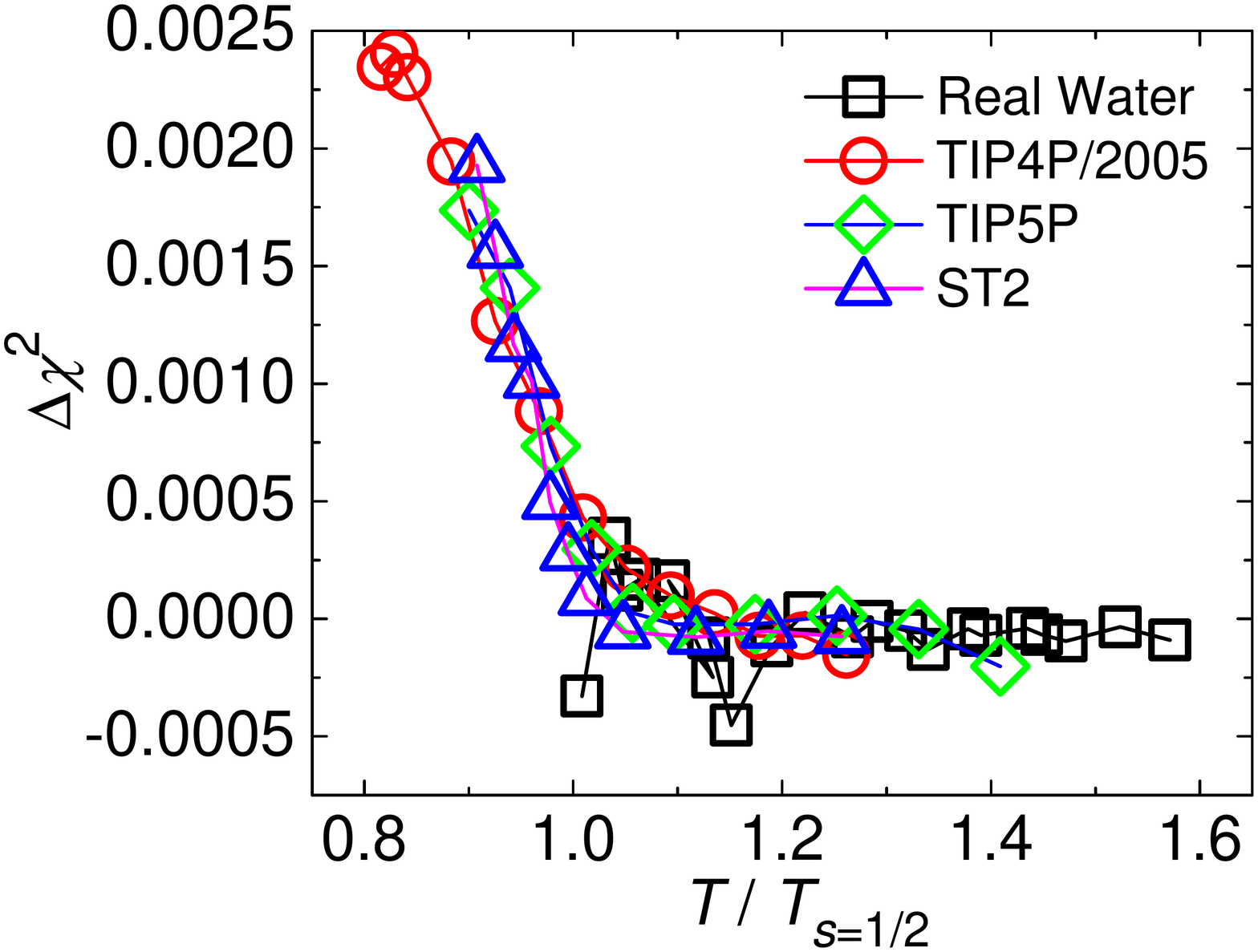}
	\end{center}
	\vspace{-8mm}
	\caption{The fitting quality of structure factors. Two schemes were applied to the analysis of the partial O-O structure factor $S_\mathrm{OO}(k)$ of liquid water. In Scheme I, the ``apparent'' first three peaks (from low to high wave number) in $S_\mathrm{OO}(k)$ are fitted to one Gaussian (1st peak), one Lorentizan (2nd peak), and one Gaussian (3rd peak) function, respectively. All the fitting parameters are free, and thus we need a new set of 9 parameters for each temperature. In Scheme II, the ``apparent'' first three peaks (from low to high wave number) in $S_\mathrm{OO}(k)$ are fitted to one Lorentzian + one Gaussian (1st peak), one Lorentizan (2nd peak), and one Gaussian (3rd peak) function, respectively. We constrain the temperature dependence of the fitting parameters. We need only up to 25 free fitting parameters for simultaneous fitting of  $S_\mathrm{OO}(k)$ at all the temperatures (see Methods). Scheme I is chosen for comparison, since it reasonably well describes $S_\mathrm{OO}(k)$ of liquid water at high temperatures with only three functions (8 fitting parameters at each temperature with position of $k_\mathrm{T3}$ peak being fixed). The mean squared residual $\chi^2$ measures the deviation of the fits from the data. The difference in the mean squared residual between Schemes I and II, $\Delta \chi^2 = \chi^2_\mathrm{I} - \chi^2_\mathrm{II}$, is shown for real water, simulated TIP4P/2005, TIP5P and ST2 waters as a function of scaled temperature $T/T_{s=1/2}$. 
		All the curves collapse into a master curve, suggesting a universal behavior independent of models. For all the systems, $\Delta \chi^2$ remains almost zero above $1.1T_{s=1/2}$, whereas sharply increases below $1.1T_{s=1/2}$, reflecting the fast growth of LFTS upon cooling (Fig.~4a). The failure of Scheme I at lower temperatures strongly suggests that the structure of liquid water fundamentally changes with temperature. At low enough temperatures (below $\sim 1.1T_{s=1/2}$), liquid water can no longer be treated to be microscopically homogeneous unlike at high temperatures, and should be regarded as a dynamical mixture of two local structure motifs. The two structure motifs have different contributions to the ``apparent'' first diffraction peak in $S_\mathrm{OO}(k)$ (Figs.~2-3), leading to the failure of Scheme I that assumes only a single type of contribution to the peak. For real water, there are three data points significantly deviating from the master curve, which are attributed to the experimental errors in the structure factor data (see Figure~\ref{fig:noise})}
	\label{fig:fitError}
\end{figure}

\begin{figure}[h!]
	\begin{center}
		\includegraphics[width=16cm]{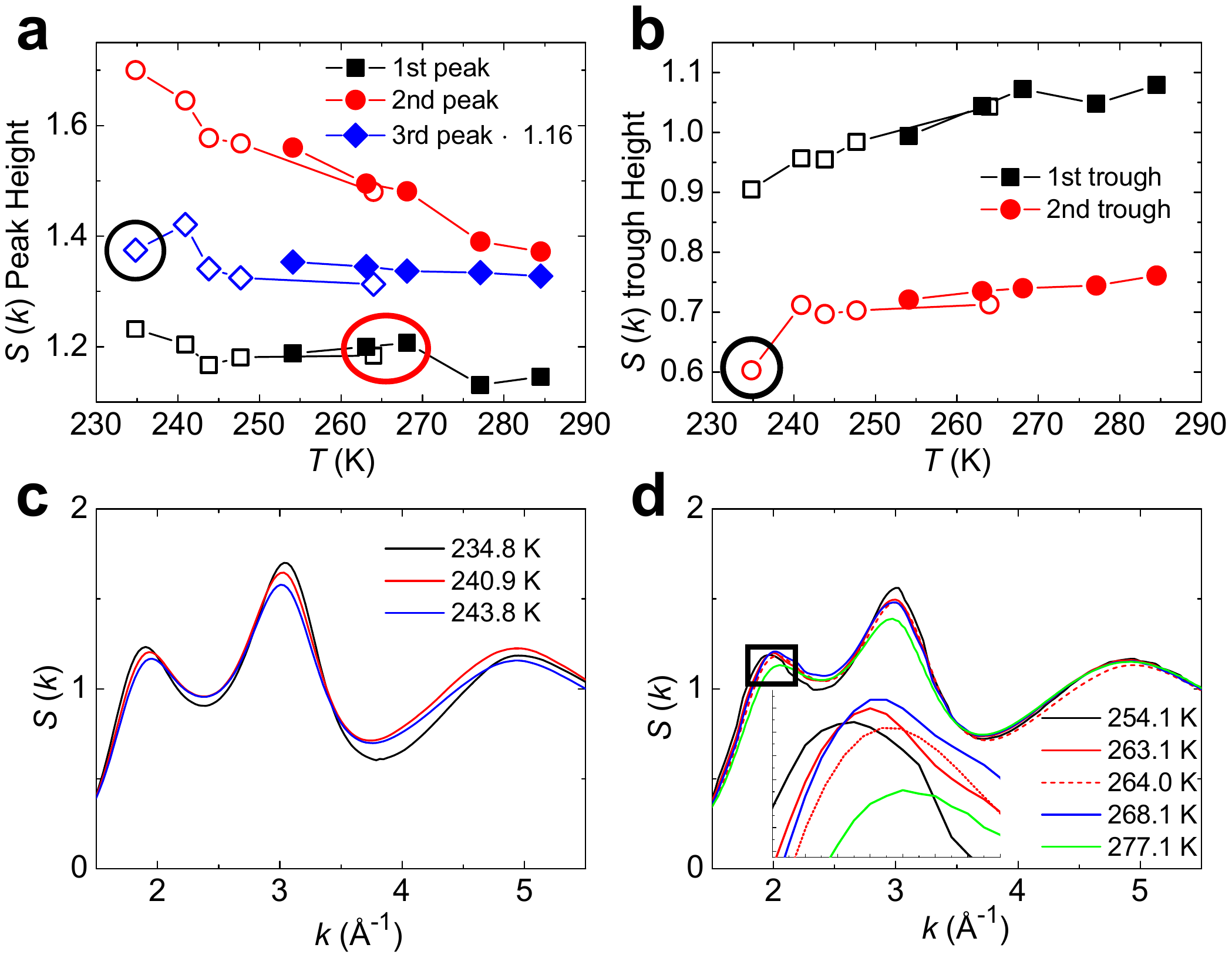}
	\end{center}
	\caption{Noise analysis in the experimental structural factor of supercooled water. {\bf a}, The height of the 1st, 2nd and 3rd peak in the experimental partial O-O structure factor as a function of temperature. The height of the 3rd peak is multiplied by a factor of 1.16 for clarification. {\bf b}, The height of the 1st and 2nd trough in the experimental partial O-O structure factor as a function of temperature. In {\bf a} and {\bf b} the filled and open symbols are data from Ref.~\cite{skinner2014structure} and~\cite{pathak2019intermediate}, respectively. {\bf c}, The experimental partial O-O structure factor of supercooled water at 234.8, 240.9 and 243.8~K (Ref.~\cite{pathak2019intermediate}). {\bf d}, The experimental partial O-O structure factor of supercooled water at 254.1, 263.1, 264.0, 268.1 and 277.1~K (solid line from Ref.~\cite{skinner2014structure} and dashed line from Ref.~\cite{pathak2019intermediate}). The inset shows the enlarged plot of the 1st peak inside the black square. Clearly, the structure factors measured at 234.8 K, 264.0 K and 268.1 K suffer from large uncertainty, because they significantly deviate from the normal temperature dependence, as highlighted by the black and red circles in {\bf a} and {\bf b}. The large errors at these three temperatures result in the three outliers in the $\Delta \chi^2$ value for real water, deviating from the master curve. For example, at 234.8 K the 3rd peak and the 2nd trough of the structure factor is substantially smaller than the expectation from the temperature trend, which is the major source of the deviation of the fits from the data by our fitting scheme II (see Figure~\ref{fig:sqoofit}a). We note that scheme I independently fits structure factors at different temperatures, whereas scheme II simultaneously fits structure factors at all the temperatures. Because scheme II uses much less fitting parameters, it is more robust, but at the same time, more sensitive to the errors in the data than scheme I.}
	\label{fig:noise}
\end{figure}

\begin{figure}[h!]
	\begin{center}
		\includegraphics[width=10cm]{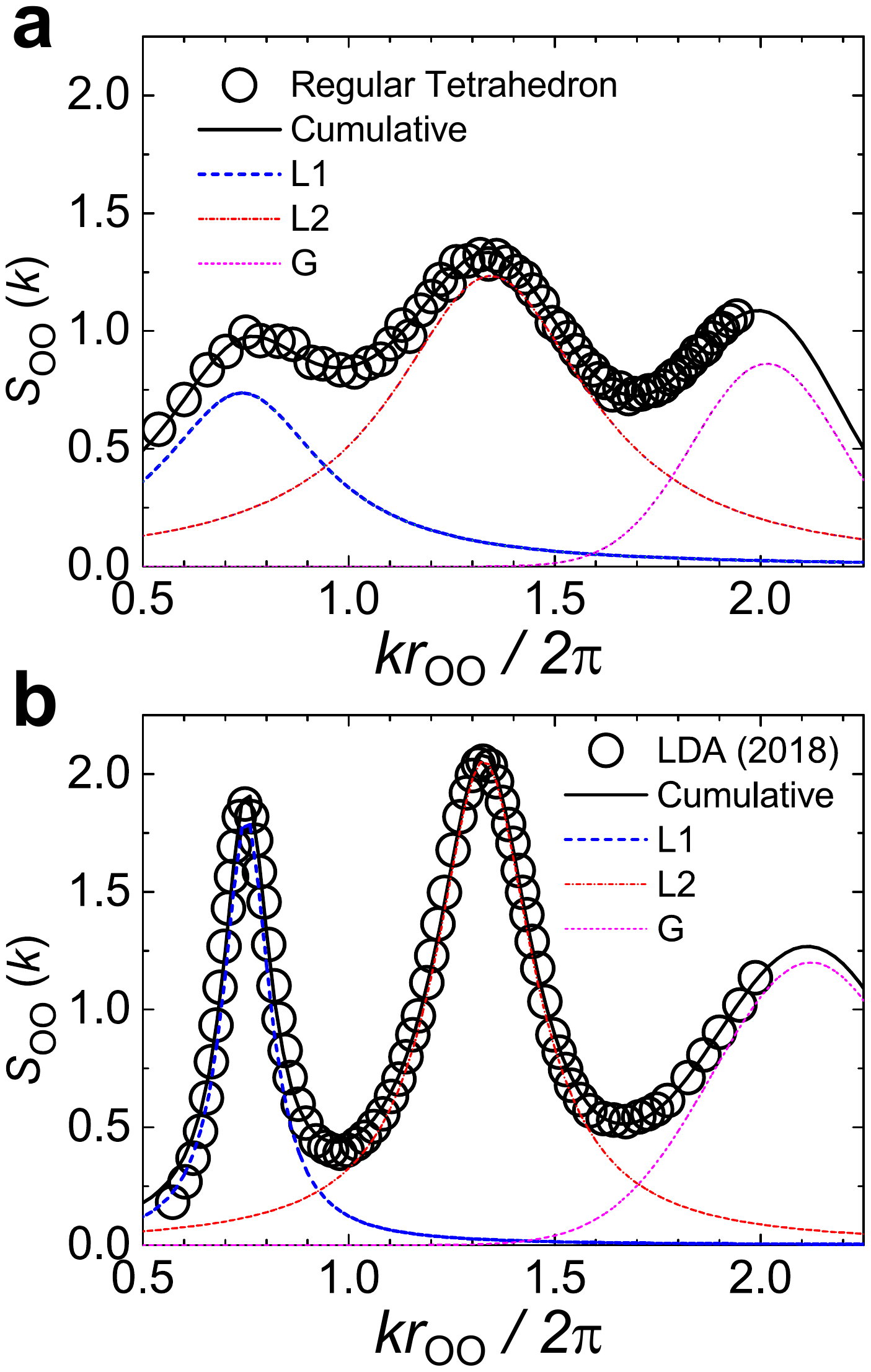}
	\end{center}
	\caption{Analysis of the O-O partial structure factor of a single regular tetrahedron and experimental LDA ice. {\bf a,} Decomposition of the structure factor of a single regular tetrahedron (Fig.~2b) by two Lorentzian (L1 + L2) and one Gaussian (G) functions. {\bf b,} Decomposition of the structure factor of experimental LDA ice~\cite{mariedahl2018} by the same functions. The three characteristic peaks obtained from the fit are displayed by blue, orange and magenta curves. The coherence lengths of a single regular tetrahedron and experimental LDA ice were thus estimated to be $\sim2$~\AA\,\ and $\sim6.5$~\AA\,\, respectively, from the widths of the corresponding FSDP's (blue curves). The wave number is scaled by the nearest neighbor O-O distance $r_\mathrm{OO}$.}
	\label{fig:sqooTetraLDA}
\end{figure}

\clearpage




\end{document}